\documentclass[letterpaper,english,reprint,aps,superscriptaddress]{revtex4-2}

\usepackage{amsmath}
\usepackage{amssymb}
\usepackage{soul}
\usepackage{float}
\usepackage{graphicx}
\usepackage{subfig}
\usepackage{colortbl}
\usepackage[colorlinks=true,citecolor=blue,linkcolor=blue]{hyperref}
\usepackage{tikz}
\usepackage[ruled,vlined]{algorithm2e}
\usepackage{booktabs}

\newcommand{\ket}[1]{\ensuremath{\left|#1\right\rangle}} 
\newcommand{\bra}[1]{\ensuremath{\left\langle#1\right|}}

\begin{document}

\title{Natural Evolutionary Strategies for Variational Quantum Computation}

\author{Abhinav Anand}
\email[E-mail:]{abhinav.anand@mail.utoronto.ca}
\affiliation{Chemical Physics Theory Group, Department of Chemistry, University of Toronto, 80  St, George Street, Toronto, Ontario M5S 3H6 Canada.}
\author{Matthias Degroote}
\affiliation{Chemical Physics Theory Group, Department of Chemistry, University of Toronto, 80  St, George Street, Toronto, Ontario M5S 3H6 Canada.}
\affiliation{Department of Computer Science, University of Toronto, 40 St. George Street, Toronto, Ontario, M5S 2E4, Canada.}
\author{Al\'{a}n Aspuru-Guzik}
\email[E-mail:]{aspuru@utoronto.ca}
\affiliation{Chemical Physics Theory Group, Department of Chemistry, University of Toronto, 80  St, George Street, Toronto, Ontario M5S 3H6 Canada.}
\affiliation{Department of Computer Science, University of Toronto, 40 St. George Street, Toronto, Ontario, M5S 2E4, Canada.}
\affiliation{CIFAR AI chair, Vector Institute, 661 University Ave. Suite 710, Toronto, Ontario, M5G  1M1, Canada.}
\affiliation{Canadian Institute for Advanced Research (CIFAR) Lebovic  Fellow, 661 University Ave, Toronto, ON M5G 1M1, Canada.}

\begin{abstract}
Natural evolutionary strategies (NES) are a family of gradient-free black-box optimization algorithms.
This study illustrates their use for the optimization of randomly-initialized parameterized quantum circuits (PQCs) in the region of vanishing gradients.
We show that using the NES gradient estimator the exponential decrease in variance can be alleviated. 
We implement two specific approaches, the exponential and separable natural evolutionary strategies, for parameter optimization of PQCs and compare them against standard gradient descent.
We apply them to two different problems of ground state energy estimation using variational quantum eigensolver (VQE) and state preparation with circuits of varying depth and length.
We also introduce batch optimization for circuits with larger depth to extend the use of evolutionary strategies to a larger number of parameters. 
We achieve accuracy comparable to state-of-the-art optimization techniques in all the above cases with a lower number of circuit evaluations.
Our empirical results indicate that one can use NES as a hybrid tool in tandem with other gradient-based methods for optimization of deep quantum circuits in regions with vanishing gradients.
\end{abstract}

\maketitle

\section{Introduction}\label{sec:intro}
Rapid developments in quantum hardware~\cite{arute2019quantum, jurcevic2020demonstration, pino2020demonstration} have led to the development of a wide variety of algorithms to perform useful tasks on the available devices. The limited coherence times and number of qubits of the devices places limitations on the circuits that can be reliably executed. A class of algorithms which are well-suited for these noisy intermediate scale quantum (NISQ)~\cite{preskill2018quantum} devices are the variational quantum algorithms (VQAs)~\cite{mcclean2016theory}. These algorithms use classical and quantum computers in tandem, using the quantum computers for state preparation with parametrized unitaries and measurement, and then optimizing the parameters of the unitaries on a classical computer in an iterative manner. 

VQAs have been used for a variety of tasks such as simulating molecules and many body systems~\cite{peruzzo2014variational,cao2019quantum,mcardle2020quantum}, compression of quantum states~\cite{romero2017quantum,wan2017quantum,pepper2019experimental}, classification tasks~\cite{schuld2020circuit,grant2018hierarchical,PerezSalinas2020datareuploading}, generative modelling~\cite{romero2019variational,zhu2019training,lloyd2018quantum,dallaire2018quantum,hu2019quantum, anand2020experimental},  and solving combinatorial optimization problems~\cite{farhi2014quantum} among others.~\cite{kottmann2020quantum,kyaw2020quantum,bharti2021noisy}
However, these algorithms suffer in cases when there is little or no knowledge about a good initial point. For such problems, the optimization of the unitaries using gradient-based methods become impossible for deep quantum circuits, as the average and variance of the gradient of the objective function decay exponentially as a function of the number of qubits and layer in the circuits~\cite{mcclean2018barren}. More recently it was reported~\cite{cerezo2020cost,wang2020noise,marrero2020entanglement} that the barren plateaus are not only present in deep circuits, but are dependent on the cost function used for optimization and that optimization of problems with global cost function is not possible for randomly parameterized shallow circuits either. Furthermore, another study~\cite{uvarov2020on} showed that the variance of the gradient is bounded by the width of the causal cone of the circuit corresponding to the various Pauli terms in the decomposition of the cost function. In another recent study~\cite{arrasmith2020effect}, the authors investigated the use of gradient-free optimizers in a barren plateau region of the landscape, and show that it becomes increasing difficult to optimize as the number of shots required grows exponentially.
 
Recently, there have been different studies that have tried to solve the barren plateau problem. One approach circumvents the problem by formulating better initial guesses for the values of the parameters~\cite{verdon2019learning}. A different study~\cite{volkoff2020large} investigated ways to boost the gradients for certain problems by using correlated parameters. In this study we use an approach based on natural evolution strategies and batch optimization for avoiding barren plateaus in the optimization landscapes.  

Evolutionary Strategies (ES)~\cite{rechenberg1978evolutionsstrategien, schwefel1977numerische} have been applied extensively in classical machine learning tasks as a viable black-box optimization tool for high-dimensional problems.
These strategies are promising for optimization of quantum circuits with respect to both the number of function evaluations and the scaling with the size of the circuit. 
The gradient is estimated by a number of function evaluations that does not increase with the number of parameters. Moreover, these function evaluations are fully independent and could be executed in parallel.
This means that these methods are suitable for high-dimensional problems.
The optimization of parametrized quantum circuits in quantum computing is an example of such high-dimensional problem.
The application of these optimization strategies to variational quantum algorithms and the adaptation to the specific challenges arise in this context pose an interesting research question.
They can be further improved by using estimates of natural search gradients, fitness shaping, and adaptive sampling~\cite{wierstra2008natural,glasmachers2010exponential} for both highly correlated (considering the full co-variance matrix) and separable (only considering a diagonal co-variance matrix) problems.
In a recent work~\cite{zhao2020natural}, the authors illustrated the use of natural evolution strategies (NES) for approximate combinatorial optimization problems, and show that they can achieve state-of-the-art performance. 
More recently in a work~\cite{coopmans2020protocol} NES was used to investigate the problem of manipulating Majorana bound-states in a topological superconductor, where the authors show that these strategies can be used efficiently to study quantum mechanical many-body systems.
In this study we use two modified strategies, separable NES and exponential NES~\cite{glasmachers2010exponential, wierstra2014natural}, for optimizing randomly initialized PQCs by considering circuit parameters as multivariate normal distributions.
We show that optimization of randomly initialized parameterised quantum circuits can be performed with significantly less circuit evaluations using NES, while achieving comparable accuracy to gradient-based methods. We also show that they can be used to amplify gradients and improve parameter initialization for gradient-based approaches. 

The rest of the paper is organised as follows, we provide the theory behind NES in section~\ref{sec:nes}, the details of the numerical experiments in section~\ref{sec:application}, the result from optimization of randomly parameterized quantum circuits in section~\ref{sec:simulations}, and finally present some concluding remarks in section~\ref{sec:conclusion}.

\section{Natural Evolution Strategies}\label{sec:nes}

Evolution strategies were first introduced by Ingo Rechenberg~\cite{rechenberg1978evolutionsstrategien} and Hans-Paul Schwefel~\cite{schwefel1977numerische}, to cope with high-dimensional continuous-valued domains.
It has remained an active field of research, and has been developed extensively over the years to include the full covariance matrix~\cite{hansen2001completely} for efficient optimization.
In what follows, we describe a specific subclass of evolution strategies called the natural evolution strategy (NES) for iteratively updating a search distribution by ascending the surface along estimated search gradients. 
The reader is redirected to Ref. \cite{wierstra2014natural, wierstra2008natural} for a detailed description of NES.

\subsection{Canonical Natural Evolution Strategies}
The NES uses search gradients to iteratively update the distribution of a stochastic variable $\boldsymbol{z}$. 
The search gradient can be defined for any distribution for which the derivatives of the log-density can be calculated. 
If we consider a distribution with density $\pi(\boldsymbol{z}|\theta)$ with $\theta$ as the parameter and $f(\boldsymbol{z})$ as the fitness for $\boldsymbol{z}$ sampled from the distribution, then we can write the expected fitness as
\begin{equation}
	J(\theta) = E_{\theta}[f(\boldsymbol{z})] = \int f(\boldsymbol{z})\pi(\boldsymbol{z}|\theta)\mathrm{d}\boldsymbol{z}.
\end{equation}
Using the 'log-likelihood trick' we can then approximate the search gradient from samples $\boldsymbol{z}_1...\boldsymbol{z}_k$ as
\begin{equation}
    \nabla_{\theta}J(\theta) \approx \frac{1}{k} \sum_{n=1}^k f(\boldsymbol{z}_n) \nabla_{\theta} \log \pi(\boldsymbol{z}_n|\theta).
\end{equation}
where $k$ is the population size. This equation has the benefit that it does not require derivatives of the fitness function. One can estimate the gradient for different commonly used search distributions including Gaussian and Cauchy distributions, and use standard gradient descent to iteratively update the parameters of the distribution
\begin{equation}
    \theta = \theta + \eta \nabla_{\theta}J(\theta),
\end{equation}
where $\eta$ is the learning rate. 

For a Gaussian prior distribution on the perturbation~\cite{salimans2017}, the procedure is shown in Algorithm~\ref{al:nes}\\ 

\begin{algorithm}[H]\label{al:nes}
\SetAlgoLined
	\textbf{input}: $f$, $\boldsymbol{\mu}_{init}$, $\sigma_{init}$ \\
 \While{stopping criterion is not met}{
  \For{n = 1..k}{
   draw sample $s_n \sim ~N(0,I)$\\
	$\boldsymbol{z}_n = \boldsymbol{\mu} + \sigma_{init}\boldsymbol{s}_n$\\
	
   evaluate the fitness $f(\boldsymbol{z}_n)$
   }{
   compute gradients: \\
	$\nabla J = \frac{1}{k}\sum_{n=1}^k f(\boldsymbol{z}_n)\cdot \boldsymbol{s}_n$\\
   update parameters: \\
	$\boldsymbol{\mu} = \boldsymbol{\mu} + \eta \cdot \frac{1}{\sigma_{init}} \nabla J$ \\
  }
}
 \caption{Canonical ES}
\end{algorithm} 
\medskip
where $f$ is the fitness function, $\eta$ is the learning rate, $\sigma_{init}$ is the fixed variance, $\boldsymbol{s}_n$ is a Gaussian perturbation sampled from the normal distribution $N(0,I)$.
This procedure replaces the normal gradient with the evaluation of the objective function blurred with Gaussian noise and samples the stochastic gradient with the perturbed parameters.
It reduces to the simultaneous perturbation stochastic approximation (SPSA)~\cite{Spall1992} for the case of a symmetric perturbation forward and backward for a single Gaussian perturbation.
As this method does not follow the exact gradient of the cost function but a stochastic estimator, it does not necessarily get stuck on the same barren plateaus as the analytical gradient.
We will show in the numerical section that this is the case in all the simulations we performed.

Instead of using the standard gradients for updates, modern evolutionary strategies implementations use the natural gradients because of their numerous advantages~\cite{amari1998natural,amari1998natural2}.
The natural gradient can be calculated using the Fisher matrix $\boldsymbol{F}$ which can be estimated from the search gradient as
\begin{equation}
    \boldsymbol{F} \approx \frac{1}{k} \sum_{n=1}^k \nabla_{\theta}\log\pi(\boldsymbol{z}_n|\theta)  \nabla_{\theta}\log\pi(\boldsymbol{z}_n|\theta)^T,
\end{equation}
and the distribution is now updated as
\begin{equation}
    \theta = \theta + \eta \cdot \boldsymbol{F}^{-1}\nabla_{\theta}J(\theta).
\end{equation}
The canonical NES can be further improved by fitness shaping and adaptive sampling, which are used in exponential Natural Evolution Strategies (xNES) and separable Natural Evolution Strategies (sNES)~\cite{wierstra2008natural,glasmachers2010exponential,wierstra2014natural}.

\subsection{Exponential Natural Evolution Strategies}
The multivariate Gaussian distribution is one of the most prominent search distribution used for evolutionary strategies.
Historically this distribution has received the most attention because of its analytical properties and the resulting elegance of the equations. 
Based on these techniques a new algorithm known as Exponential Natural Evolution Strategies (xNES) can be formulated, for which we provide the outline in Algorithm~\ref{al:xnes}, as presented in Ref.~\cite{wierstra2014natural,glasmachers2010exponential}.
It goes beyond the standard NES by introducing an exponential parametrization by which it makes the natural gradient easier to compute.\\

\begin{algorithm}[H]\label{al:xnes}
\SetAlgoLined
\textbf{input}: $f$, $\boldsymbol{\mu}_{init}$, $\boldsymbol{\sum}_{init}$ = \textbf{AA}$^T$ \\
\textbf{initialize}: \\
$\sigma$ = $\sqrt[d]{|det(\boldsymbol{A})|}$\\
\textbf{B} = \textbf{A}/$\sigma$ \\
 \While{stopping criterion is not met}{
  
  \For{n = 1..k}{
   draw sample $\boldsymbol{s}_n \sim ~N(0,I)$\\
   $\boldsymbol{z}_n = \boldsymbol{\mu} + \sigma\boldsymbol{B}^T\boldsymbol{s}_n$\\
	
   evaluate the fitness $f(\boldsymbol{z}_n)$
   }{
   sort $\{(\boldsymbol{s}_n, \boldsymbol{z}_n)\}$ with respect to $f(\boldsymbol{z}_n)$ and compute utilities $u_n$\\
   
   compute gradients: \\
    $\nabla_{\boldsymbol{\mu}} J = \sum_{n=1}^k u_n \cdot \boldsymbol{s}_n$\\
    $\nabla_{\boldsymbol{M}} J = \sum_{n=1}^k u_n \cdot (\boldsymbol{s}_n\boldsymbol{s}_n^T - I)$\\
    $\nabla_\sigma J = $tr$(\nabla_{\boldsymbol{M}} J)/d$ \\
    $\nabla_{\boldsymbol{B}} J = \nabla_{\boldsymbol{M}} J -  \nabla_\sigma J\cdot I$\\
   update parameters: \\
   $\boldsymbol{\mu} = \boldsymbol{\mu} + \eta_{\boldsymbol{\mu}} \cdot \sigma \boldsymbol{B} \cdot \nabla_{\boldsymbol{\mu}} J$ \\
   $\sigma = \sigma \cdot \exp(\eta_s / 2 \cdot \nabla_\sigma J)$ \\
   $\boldsymbol{B} = \boldsymbol{B} \cdot \exp(\eta_{\boldsymbol{B} }/ 2 \cdot \nabla_{\boldsymbol{B}} J)$
  }
 }
 \caption{xNES}
\end{algorithm} 

where, $f$ is the fitness function, $d$ is the dimension of the problem, $\boldsymbol{A}$ is the covariance matrix, $\boldsymbol{\mu}$ is the mean, $k$ is the number of walkers, $\boldsymbol{s}_n$ is the sample in local coordinates, $\boldsymbol{z}_n$ is the sample in the task coordinates, $N(0,I)$ is a normal distribution, $\eta_{\boldsymbol{B}}$ is the covariance learning rate, $\eta_{\boldsymbol{\mu}}$ is the mean learning rate, $\eta_s$ is the scale learning rate, and $u_n$ is the utility function used for fitness shaping which is defined as
\begin{equation}\label{eq:util}
    u_n = \frac{max(0, log(\frac{k}{2} + 1) - log(n))}{\sum_{j=1}^k max(0, log(\frac{k}{2} + 1) - log(j))} - \frac{1}{k}.
\end{equation}
The stopping criterion for the xNES algorithm is met when the maximum value of the covariance matrix reaches below a threshold of 10$^{-8}$. The xNES strategy works well across a wide variety of optimization problems~\cite{wierstra2014natural}, however the computational cost increases significantly with increasing variable dimension.

\subsection{Separable Natural Evolution Strategies}
For high-dimensional problems, the more suitable variant is the separable NES (sNES). Here, the correlation between the distribution variables is not considered (the covariance matrix $\boldsymbol{A}$ is forced to be diagonal), implying that the parameter distributions of the PQCs are considered independent of each other.
The sNES algorithm is expected to perform at least as well as xNES on separable problems, to perform worse than xNES for highly correlated problems and to outperform xNES for very high-dimensional problems~\cite{wierstra2014natural}. The sNES algorithm is outlined in Algorithm~\ref{al:snes}, as in Ref~\cite{wierstra2014natural}.\\

\begin{algorithm}[H]\label{al:snes}
\SetAlgoLined
\textbf{input}: $f$, $\boldsymbol{\mu}_{init}$, $\boldsymbol{\sigma}_{init}$ \\

 \While{stopping criterion is not met}{
  
  \For{n = 1..k}{
   draw sample $\boldsymbol{s}_n \sim ~ N(0,I)$\\
   $\boldsymbol{z}_n = \boldsymbol{\mu} + \boldsymbol{\sigma}\boldsymbol{s}_n$\\
   evaluate the fitness $f(\boldsymbol{z}_n)$
   }{
   sort $\{(\boldsymbol{s}_n, \boldsymbol{z}_n)\}$ with respect to $f(\boldsymbol{z}_n)$ and compute utilities $u_n$\\
   
   compute gradients: \\
    $\nabla_{\boldsymbol{\mu}} J = \sum_{n=1}^k u_n \cdot \boldsymbol{s}_n$\\
    $\nabla_{\boldsymbol{\sigma}} J = \sum_{n=1}^k u_n \cdot (\boldsymbol{s}_n^2\ - 1)$\\
    
   update parameters: \\
   $\boldsymbol{\mu} = \boldsymbol{\mu} + \eta_{\boldsymbol{\mu}} \cdot \boldsymbol{\sigma}  \cdot \nabla_\mu J$ \\
   $\boldsymbol{\sigma} = \boldsymbol{\sigma} \cdot \exp(\eta_{\boldsymbol{\sigma}} / 2 \cdot \nabla_{\boldsymbol{\sigma}} J)$
  }
 }
 \caption{sNES}
\end{algorithm}

where $f$ is the fitness function, $\boldsymbol{\mu}$ is the mean, $\boldsymbol{\sigma}$ is the standard deviation, $k$ is the number of walkers, $\boldsymbol{s}_n$ is the sample in local coordinates, $\boldsymbol{z}_n$ is the sample in the task coordinates, $N(0,I)$ is a normal distribution, $\eta_{\boldsymbol{\mu}}$ is the mean learning rate, $\eta_{\boldsymbol{\sigma}}$ is the deviation learning rate, and $u_n$ is the utility function used for fitness shaping as defined in equation~\ref{eq:util}. The stopping criterion for the sNES algorithm is met when the maximum value of the standard deviation vector reaches below a threshold of 10$^{-8}$. 

\subsection{NES for Quantum Algorithms}

A variational quantum algorithm generally consists of finding the optimal parameters $\boldsymbol{\theta^*}$ of the parameterized quantum circuit $U(\boldsymbol{\theta})$ to minimize the associated loss function.The traditional way of finding the optimal parameters include gradient-based (free) descent towards the surface minima. We consider an approach where the parameters of the circuit are sampled from a distribution, which is updated over time to achieve the global minima. This fits well within the NES framework, where one uses a search distribution to sample the parameters and update it iteratively to reach the optimum solution. The NES framework has also been shown to perform efficiently for very high-dimensional classical problems, thus can be an important tool for optimization of quantum algorithms as they deal with a high number of parameters and challenging optimization surfaces.

We carry the optimization of PQCs using NES, by considering the parameters $\boldsymbol{\theta}$ of the PQCs as centers of a search distribution $\pi(\boldsymbol{z}|\boldsymbol{\theta})$. This search distribution is then used in NES to sample different points $\boldsymbol{z}$ and estimate the gradient on the distribution parameters. The parameters of the search distribution $\boldsymbol{\theta}$ are then optimized over time by using the estimated gradient to find the optimal parameters, $\boldsymbol{\theta^*}$ for the parameterized quantum circuit. We use two specific flavors, xNES and sNES, as illustrated in Algorithms~\ref{al:xnes} and~\ref{al:snes} for the optimization as they most naturally fit the requirements for the optimization of PQCs. \\

\section{Application to Quantum Circuits}\label{sec:application}

In this section, we illustrate the use of the NES for optimization of variational quantum algorithms (VQAs). We consider two different ansatze for the problem of state preparation. The first ansatz is a randomly parameterized quantum ciruit (RPQC) with repeating parameterized and entangling layers, as 
\begin{equation}
    U(\boldsymbol{\theta}) = U(\boldsymbol{\theta}_1 ...... \boldsymbol{\theta}_L) = \prod_{l=1}^L V(\boldsymbol{\theta}_l)W_l,
\end{equation}
where $L$ is the number of layers in the circuit, $V(\boldsymbol{\theta}_l)$ = $\bigotimes_{j=1}^Q R_{lj}(\theta_{lj})$ is a tensor product of randomly chosen pauli rotation gates $R_{lj}$, $Q$ is the number of qubits in the circuit and $W_l$ is a layer of entangling CZ gates between adjacent qubits. The second ansatz is an alternate layered parameterized quantum circuit (ALPQC) which consists of two layers of $CZ$ gates between alternate nearest neighbour qubits with layers of $R_Y$ rotation gates between the entangling layers.
\begin{equation}
    U(\boldsymbol{\theta}) = U(\boldsymbol{\theta}_1 ...... \boldsymbol{\theta}_L) = \prod_{l=1}^L V(\boldsymbol{\theta}_l)W_l V^{'}(\boldsymbol{\theta}_l)W_l^{'},
\end{equation}
where $L$ is the number of layers in the circuit, $V(\boldsymbol{\theta}_n)$ = $\bigotimes_{j=1}^{Q-1} R_Y(\theta)$ and $V^{'}(\boldsymbol{\theta}_l)$ = $\bigotimes_{j=2}^Q R_Y(\theta)$ represents the layer of the rotation gates, and $W_l$ and $W_l^{'}$ represent layers of entangling CZ gates between adjacent qubits.

\begin{figure}[htbp]
\centering
    \begin{tabular}{c}
     \includegraphics[width=0.35\textwidth]{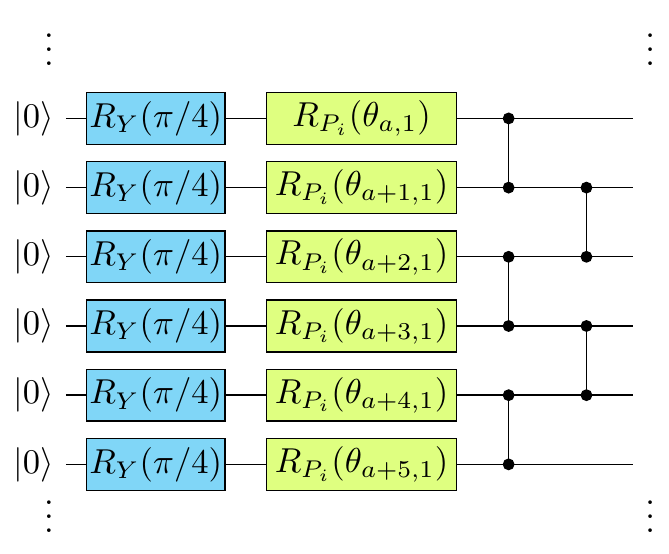}\\
     \multicolumn{1}{c}{a) A random parameterized quantum circuit.}\\
     \includegraphics[width=0.35\textwidth]{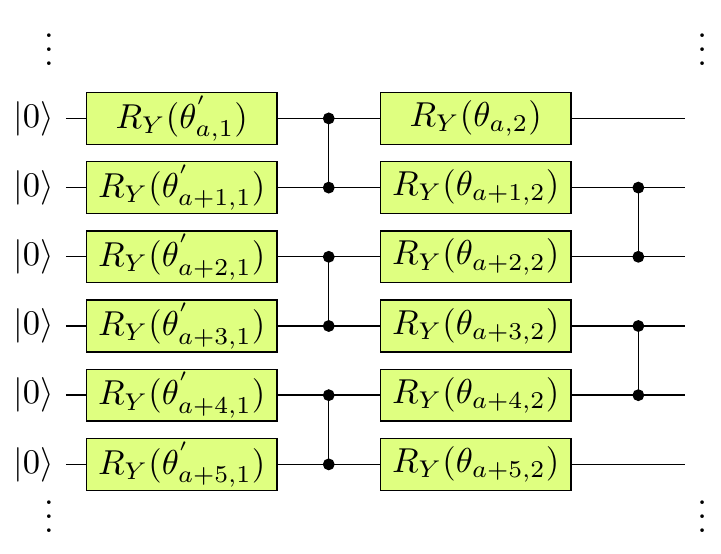}\\
	    \multicolumn{1}{c}{b)  An alternate layered parameterized circuit.}\\
	    \multicolumn{1}{c}{$\theta^{'}$ represents parameters shifted by $\frac{\pi}{4}.$}\\
     \end{tabular}
    \caption{A figure depicting a single layer of the circuits used in the different numerical simulation.}
    \label{fig:vqc_ans}
\end{figure}

The circuits are rotated out of the computational basis by adding a layer of $R_Y(\pi/4)$ gates, and are shown in Figure~\ref{fig:vqc_ans}. The objective Hamiltonian operator $H$ is chosen to be a projector $\ket{00....00}\bra{00....00}$ on the vacuum state, and the fitness function is chosen as $f(\boldsymbol{\theta}) = (1 - \bra{\psi}H\ket{\psi})^2$, where $\ket{\psi}$ = $U(\boldsymbol{\theta})\ket{00....00}$ is defined as the state after the application of the parameterized unitary on the vaccum state.

The parameters of the quantum circuits are used to define a search distribution - a multi-normal Gaussian with centers as the circuit parameters, which is then optimized using NES as discussed in section \ref{sec:nes}. 
For the xNES parameter distributions, the parameter distributions are correlated through the covariance matrix, while for the sNES the distributions are independent.
The estimates of the fitness function are carried out by calculating the average outcome of the circuit at $k$ different task coordinate values. 
The number $k$ of these evaluations, which we will reference as walkers from here on, is independent of the number of parameters and qubits in the circuit and can be fixed to a constant for a given problem. 
We perform different numerical simulations to show the application of NES for estimating the ground state energy of the water molecule in the STO-3G basis using a unitary coupled cluster (UCC) ansatz with randomly initialized parameters. We also consider a more abstract example of state preparation. In both cases we focus on the initial part of the optimization landscape to demonstrate the ability to optimize in cases with randomly initialized parameters.

\section{Numerical Simulations}\label{sec:simulations}

The circuit parameters are initialized as a multinormal distribution with randomly chosen centers and standard deviations of $0.1$. The parameters used for the optimization are given in Table~\ref{tab:params} are the same as considered in Ref.~\cite{wierstra2014natural}. These values for the different learning rates have been chosen based on the different investigations carried out in the classical machine learning literature to select the optimal value of these hyper-parameters for problems of varying dimensions~\cite{wierstra2014natural}. The value for the number of walkers $k$, has been set to 16 in most simulation after analyzing the effect of choosing a particular number is investigated in section~\ref{sec:batch_opt}. However, the value that is used widely in the machine learning community~\cite{wierstra2014natural} based on the problem dimension is $(4 + 3 \log d)$. The whole framework is implemented using \textsc{TEQUILA}~\cite{kottmann2020tequila} an open source package, which uses \textsc{QULACS}~\cite{qulacs} as the backend for the execution of all the numerical simulations.\\

\begin{table}[htbp]
    \centering
    \begin{tabular}{c|c}
        \hline
        Parameters & Values \\
        \hline
        \hline
        \vspace{2.5pt}
        $\eta_\mu$ & 1 \\ 
        \vspace{2.5pt}
        $\eta_s$ = $\eta_{\boldsymbol{B}}$ & $\frac{(9 + 3log(d))}{5d\sqrt{d}}$\\
        \vspace{2.5pt}
        $\eta_{\boldsymbol{\sigma}}$ &  $\frac{(3 + log(d))}{5d\sqrt{d}}$\\
        \hline
        \hline
    \end{tabular}
    \caption{A table containing the values of all the parameters employed for NES strategies used for optimization. $d$ represents the dimension of the batch.}
    \label{tab:params}
\end{table}

\subsection{Search Direction for NES}
\begin{figure}[htbp]
\centering
    \begin{tabular}{c}
    \toprule
	    a) $\sigma_{init} = \pi/8$ \\
    \midrule
    \includegraphics[width=0.45\textwidth]{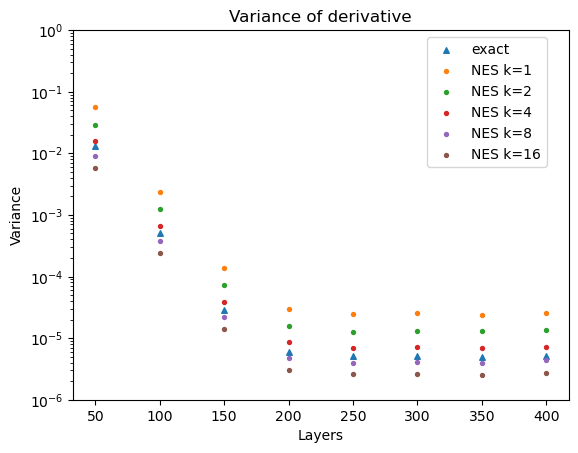} \\
    \midrule
	    b) $\sigma_{init} = \pi/16$ \\
    \midrule
    \includegraphics[width=0.45\textwidth]{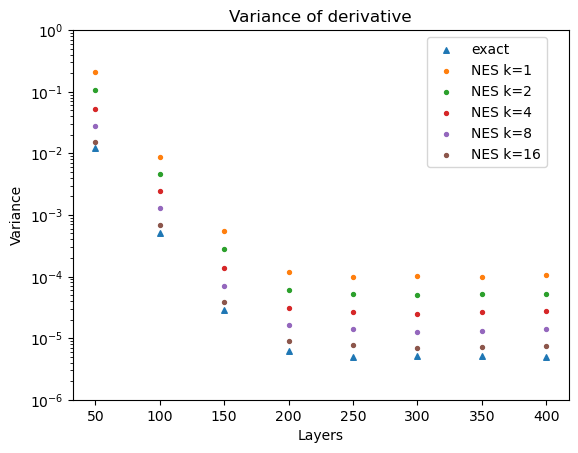} \\
    \midrule
	    c) $\sigma_{init} = \pi/32$ \\
    \midrule
    \includegraphics[width=0.45\textwidth]{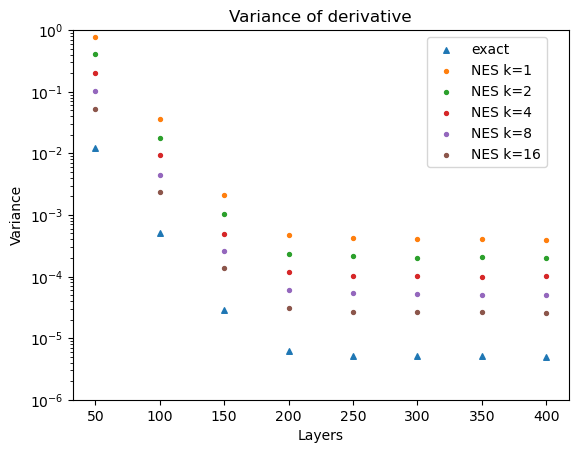} \\
    \midrule
    
    \end{tabular}
	\caption{The variance over $10000$ random initializations of the estimated search gradient with respect to the first variable for the circuit in Figure~\ref{fig:vqc_ans}(a). The plot corresponding to 'exact' represent the variance of the analytical gradient.}
    \label{fig:amplification}
\end{figure}

The problem for randomly initialized unstructured quantum circuits is that both the mean and variance tend to zero for random initializations~\cite{mcclean2018barren}. This implies that all directions to search for the optimal parameters are equally likely, thus the difficulty in optimization.
Adding random noise can not change the average value of the gradient but we want to demonstrate that we have a valid search direction in the optimization landscape. The search direction is defined by an estimate of the gradient of the search distribution on its parameters, and we show that this estimate does not have a vanishing variance. 
For this we have repeated the calculations in~\cite{mcclean2018barren} for the circuit in \ref{fig:vqc_ans}(a) for 18 qubits with a local cost function.
We average each point over $10000$ random initializations.
For each random initialization, we have calculated the estimate of the search gradient with respect to the first parameter with a varying number of walkers $k$ from $1$ to $8$.
Figure~\ref{fig:amplification} shows results of the calculations for the mean of the surrogate search gradients obtained with $\sigma_{init}$ taking values $\pi/8,\pi/16,\pi/32$.

The plots show that variance can be amplified arbitrarily as the width of the Gaussian noise is reduced.
This should not be surprising as this parameter appears in the denominator in Algorithm~\ref{al:nes}.
Decreasing the width of the noise also means that the NES samples less far around the current parameters and less information of the optimization surface is included.
As the number of walkers $k$ is increased, the stochastic approximation of the gradient approaches the true value of the gradient better and the variance is reduced.
Both the width of the noise and the number of walkers are important and need to be chosen appropriately.
In what follows we will let the width of the noise be a variable to optimize and the number of walkers a hyperparameter of our algorithm.

This section demonstrates that we can get an arbitrary amplification of the gradient for randomly initialized quantum circuits by replacing the exact gradient with a stochastic surrogate.
That does not necessarily mean that optimization would be possible following this gradient.
It is possible that our estimator leads to finding local minima or poor convergence behavior.
In the next sections we show that optimization following this gradient is possible for a wide array of problems.

\subsection{State preparation}
\begin{figure}[htbp]
\centering
    \begin{tabular}{c c}
    \toprule
    RPQCs & ALPQCs\\
    \midrule
    \multicolumn{2}{c}{\textbf{sNES}}\\
    \midrule
    \multicolumn{2}{c}{a) 5 qubits}\\
    \midrule
    \includegraphics[width=0.245\textwidth]{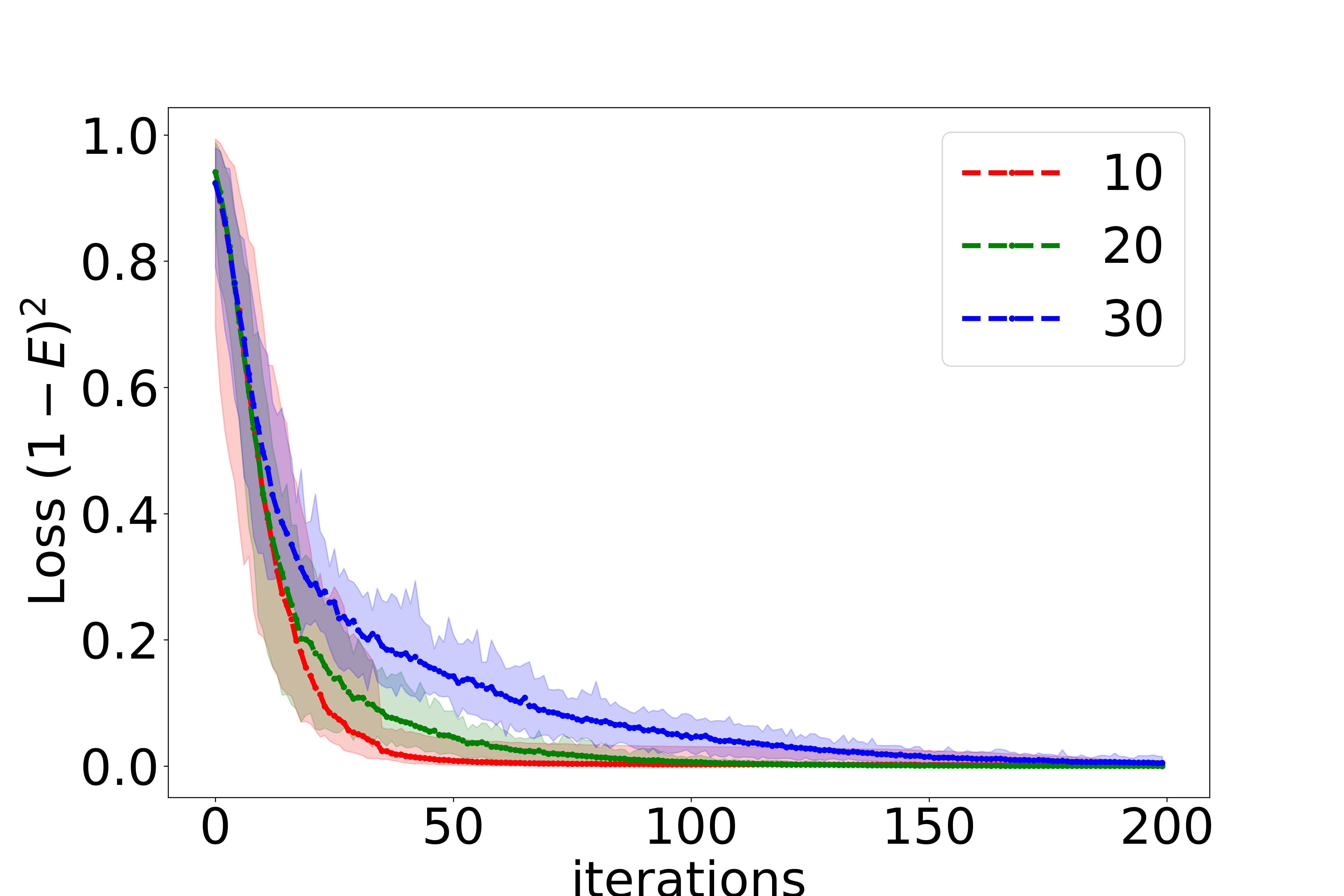} & 
     \includegraphics[width=0.245\textwidth]{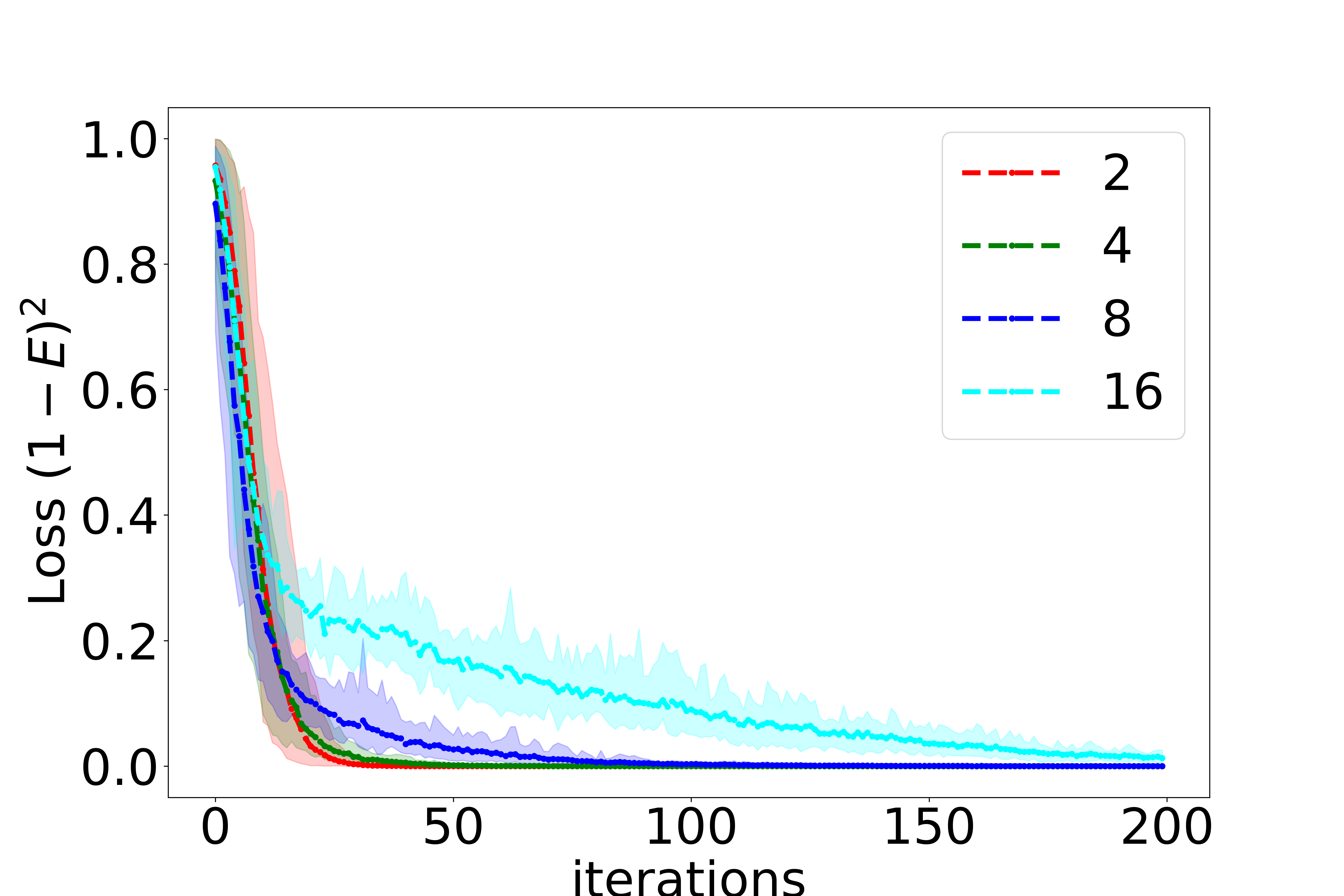}\\
     \midrule
    \multicolumn{2}{c}{b) 10 qubits }\\
    \midrule
    \includegraphics[width=0.245\textwidth]{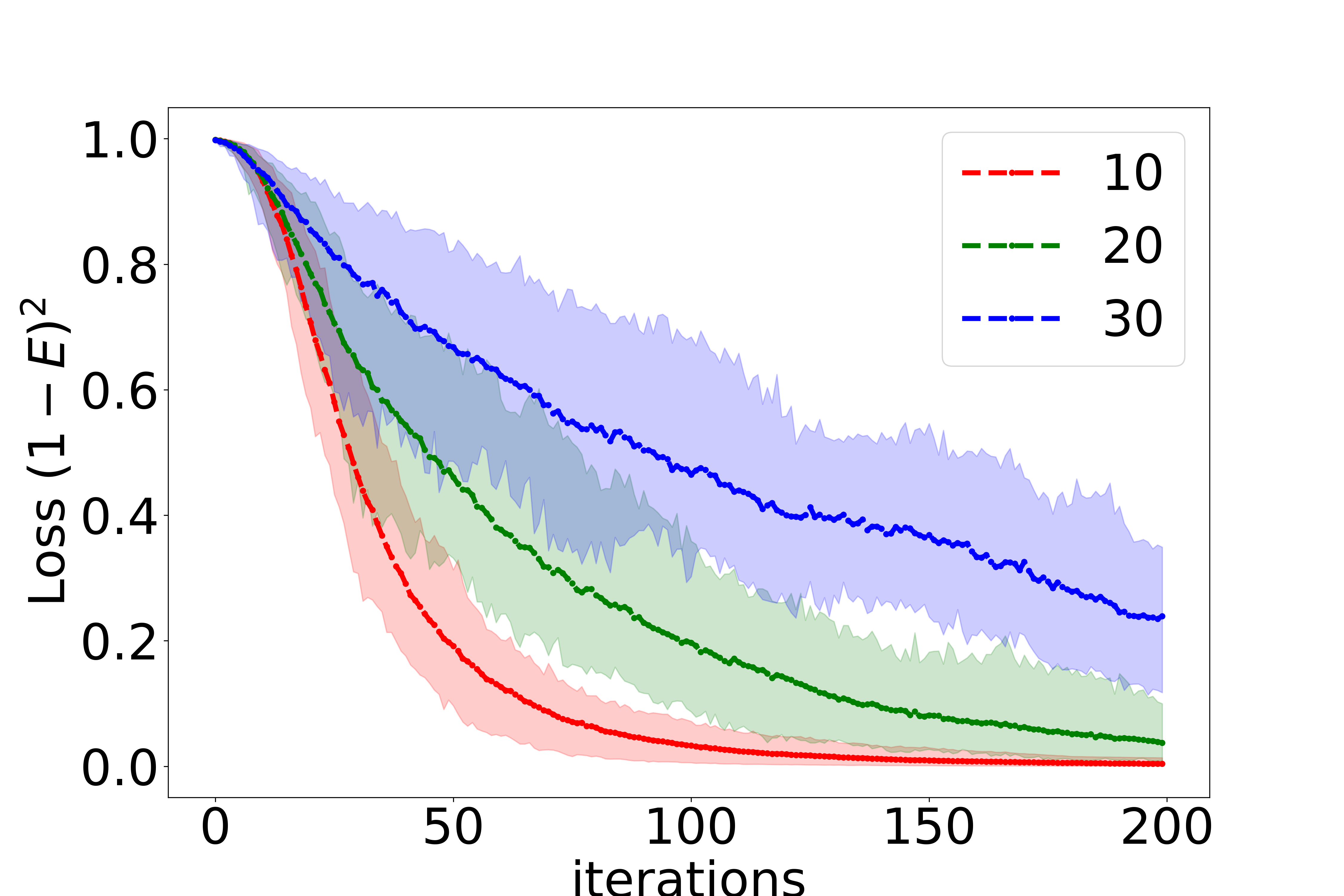} & 
     \includegraphics[width=0.245\textwidth]{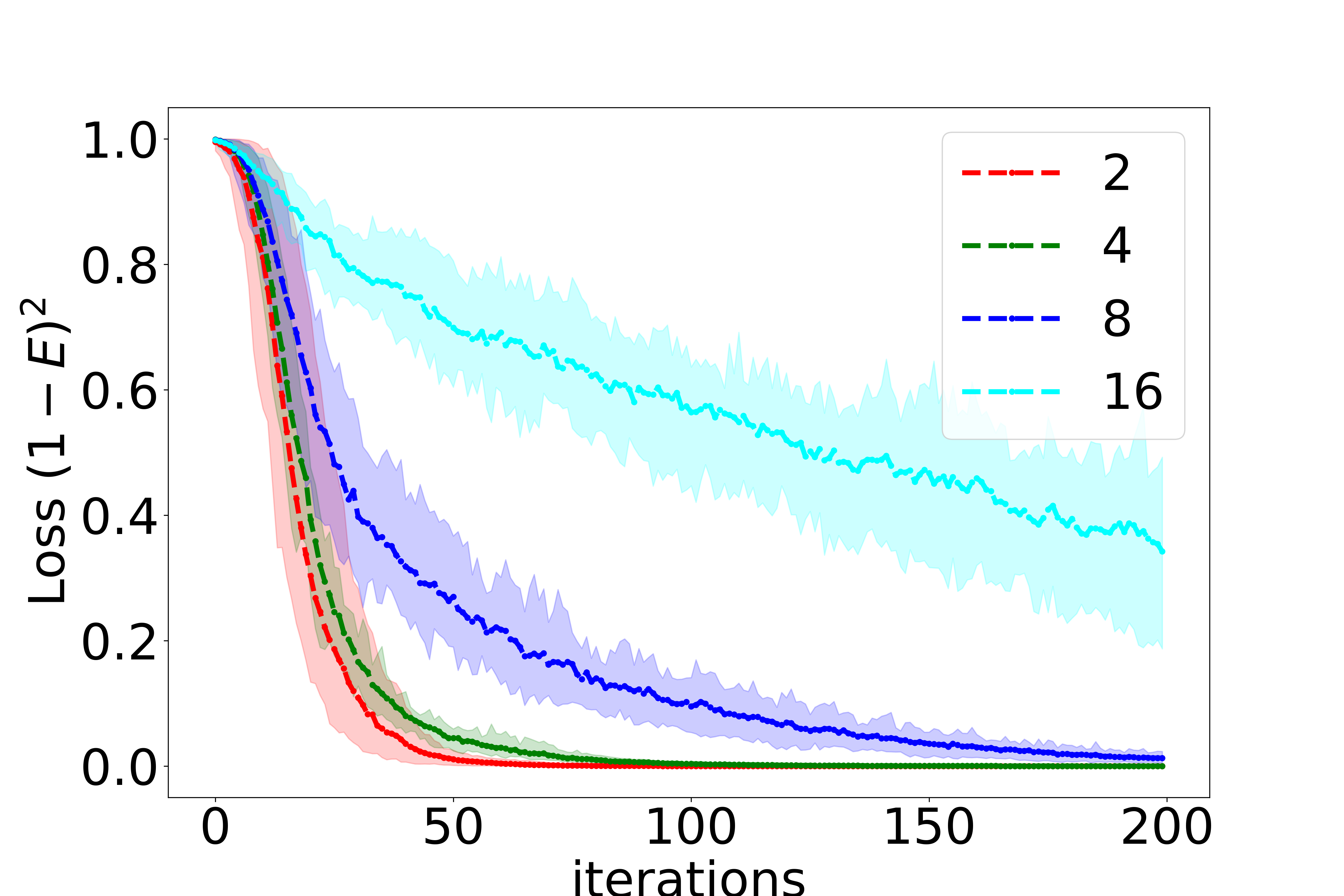}\\
     \midrule
    \multicolumn{2}{c}{c) 15 qubits}\\
    \midrule
    \includegraphics[width=0.245\textwidth]{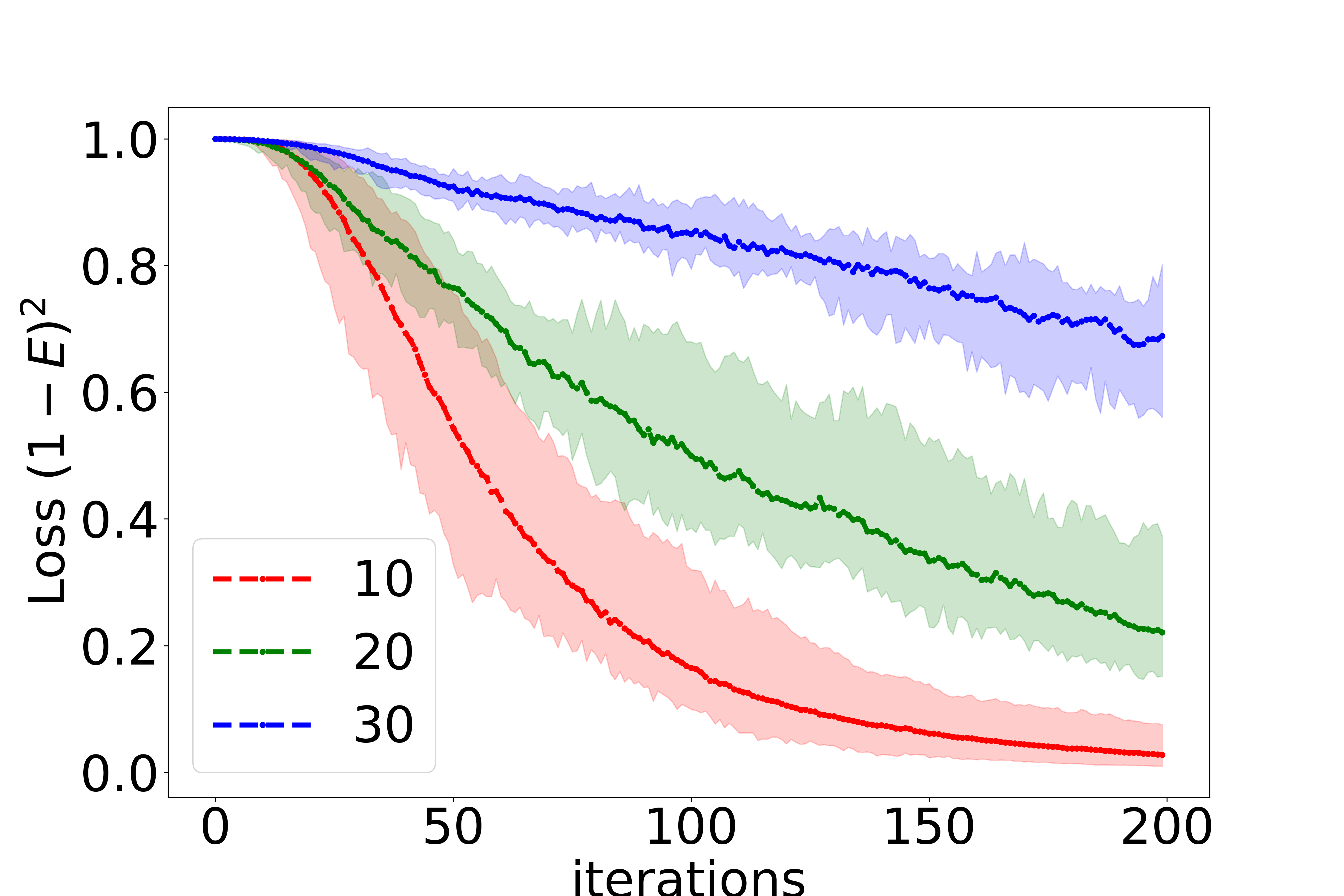} & 
     \includegraphics[width=0.245\textwidth]{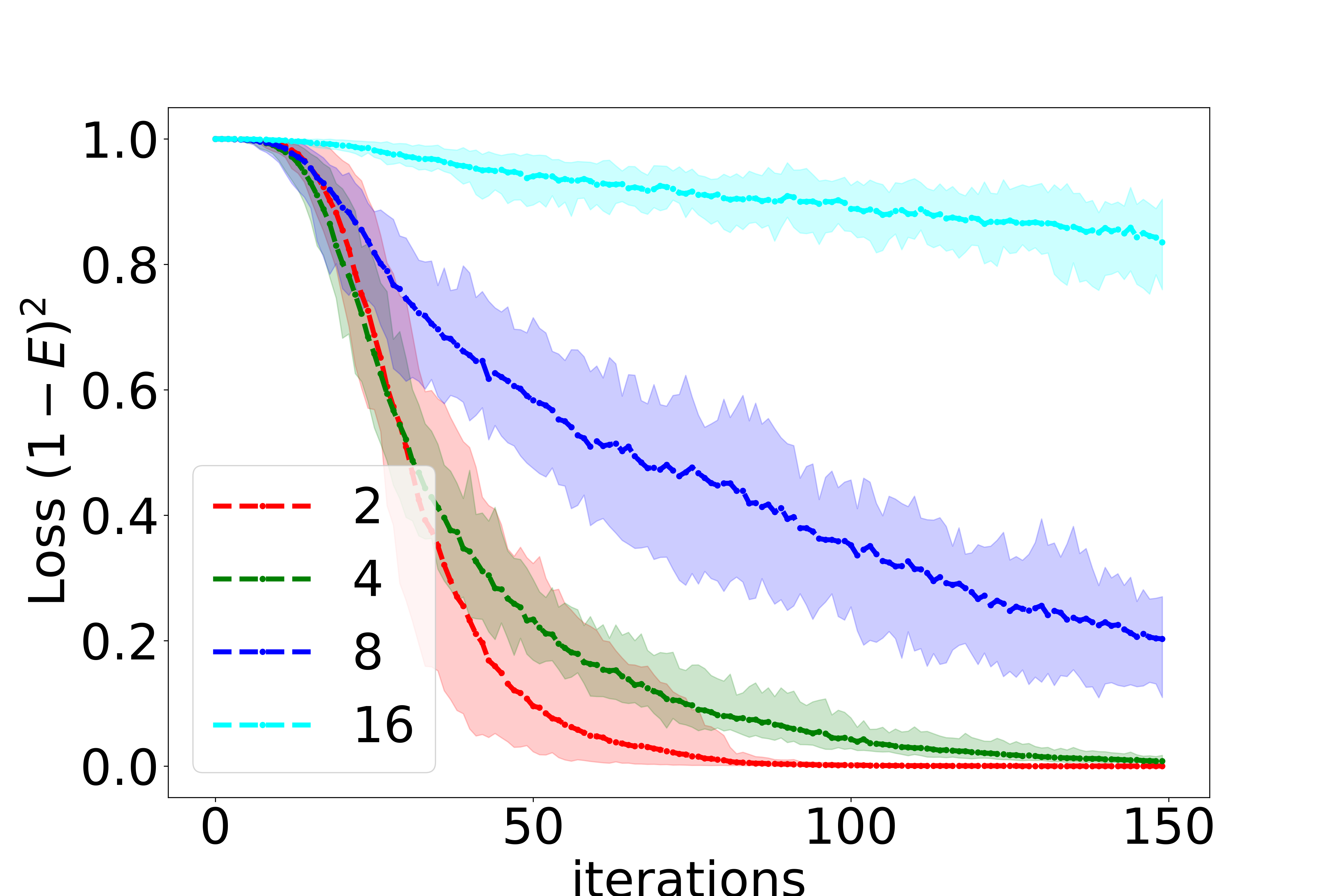}\\
     \midrule
        \multicolumn{2}{c}{\textbf{xNES}}\\
    \midrule
    \multicolumn{2}{c}{d) 5 qubits}\\
    \midrule
    \includegraphics[width=0.245\textwidth]{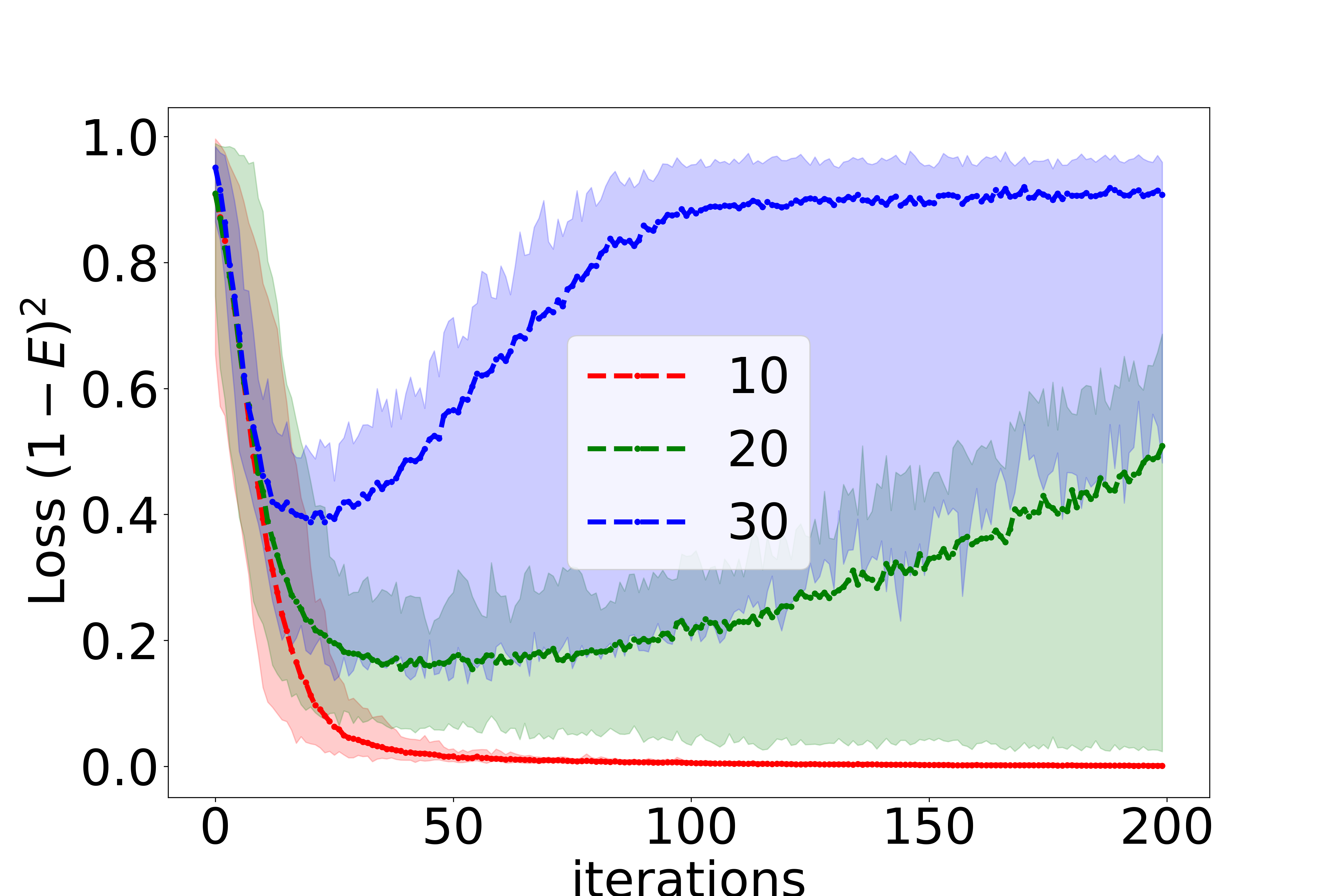} & 
     \includegraphics[width=0.245\textwidth]{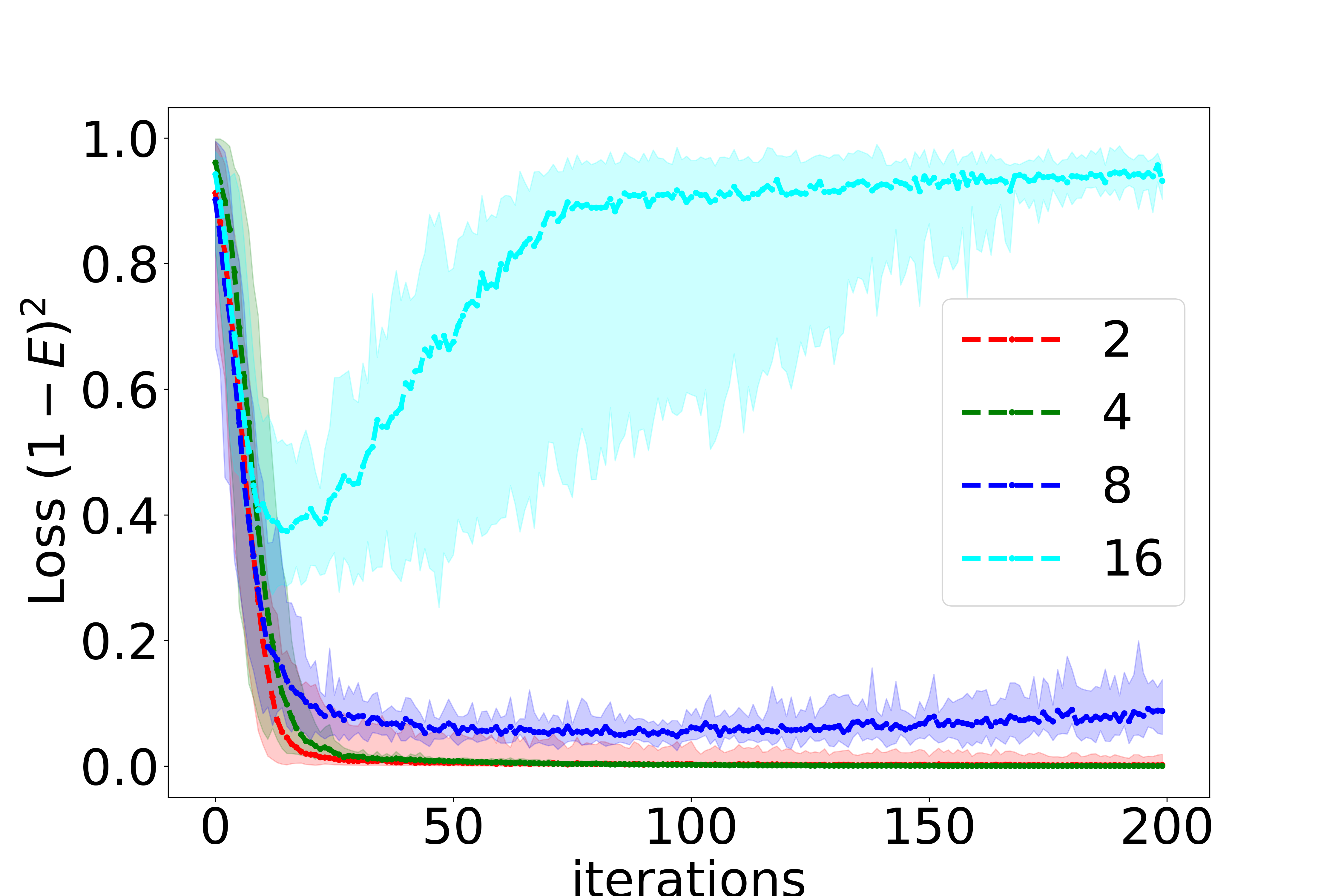}\\
     \midrule
     \multicolumn{2}{c}{e) 10 qubits}\\
    \midrule
    \includegraphics[width=0.245\textwidth]{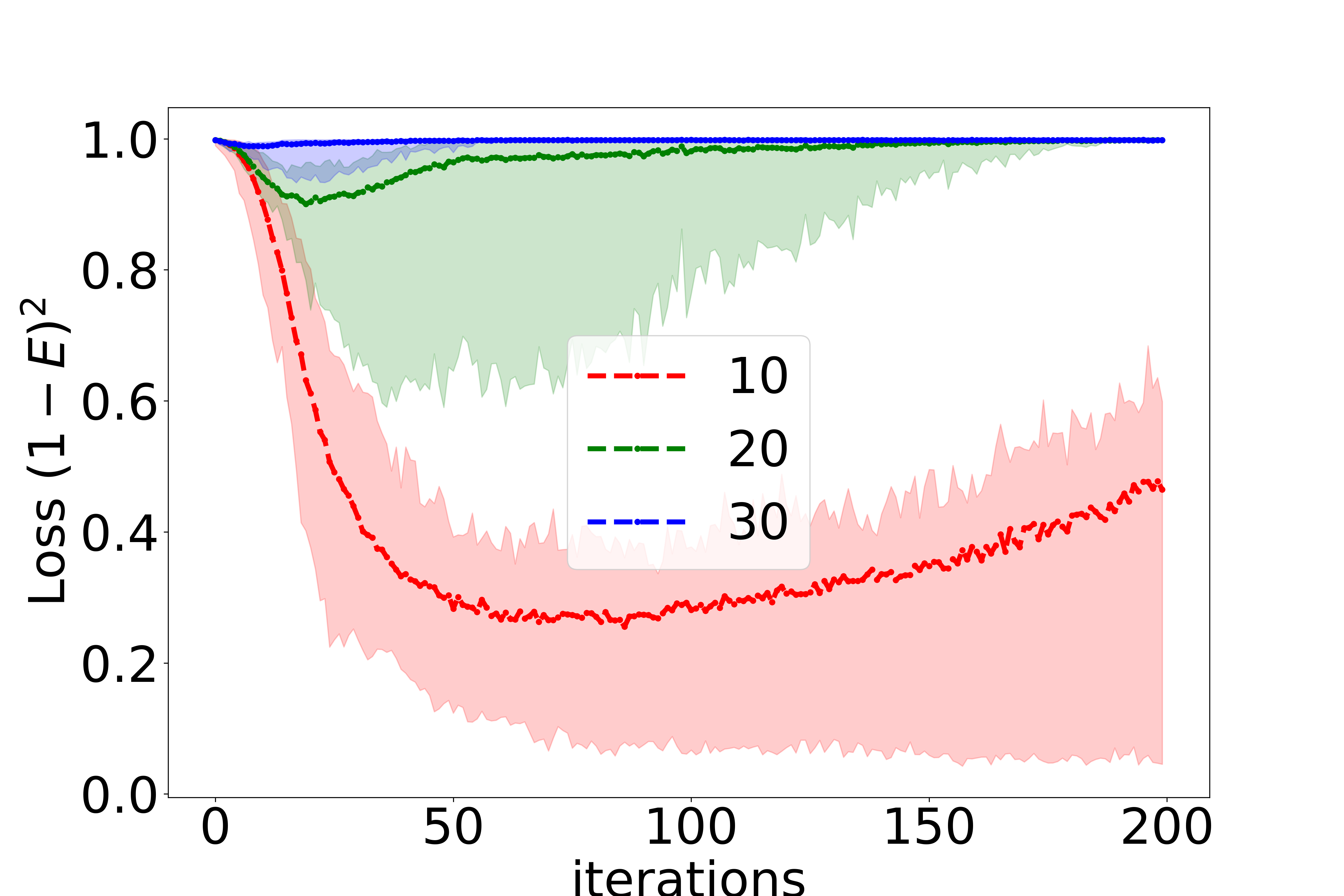} & 
     \includegraphics[width=0.245\textwidth]{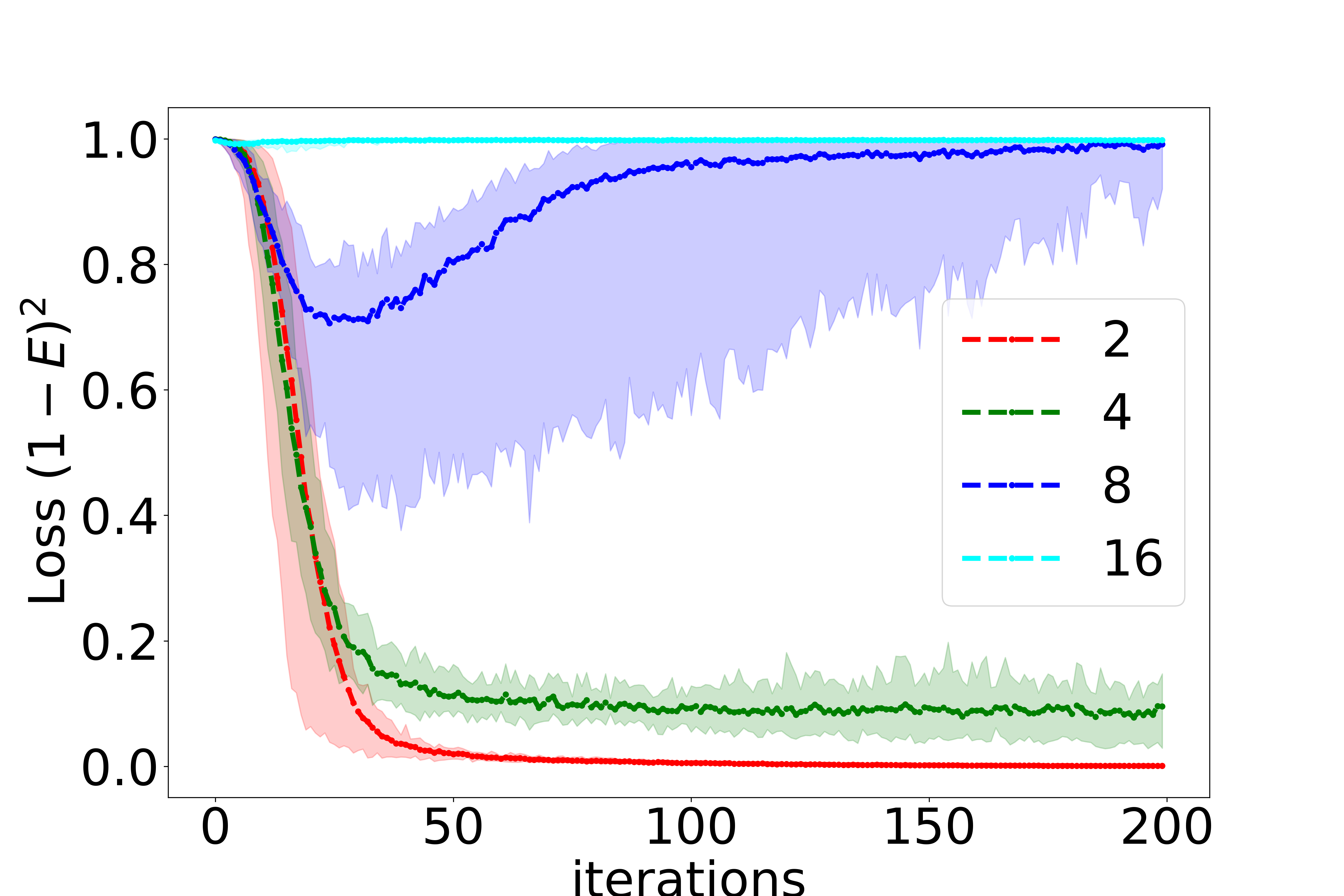}\\
     \midrule
    \end{tabular}
    \caption{Loss functions with respect to the total number of iterations for different state preparation problems of varying circuit size. The solid lines corresponds to the mean of the loss functions from 10 different runs, and the shadow represent the area between the best and worst values from the simulations. The colors in the plots correspond to different number of layers used in the parameterized quantum circuits for the problem.}
    \label{fig:full_s_nes}
\end{figure}

We carry out numerical simulations for optimization of different quantum circuits with varying number of qubits and layers to prepare the corresponding $\ket{00....00}$ state. The number of walkers for all the simulations is chosen to be 16 unless stated otherwise. We plot the results from the simulation using NES and the two different ansätze RPQCs and ALPQCs with different layers in Figure~\ref{fig:full_s_nes}.

It can be seen from the plots in Figure~\ref{fig:full_s_nes} that the optimization is possible in all cases when using sNES, but xNES fails to find the global optimum in cases with large number of parameters. 
This finding agrees with the expected behavior of sNES for these high-dimensional and multi-modal problems with highly redundant global optima, as pointed out for various cases earlier~\cite{wierstra2014natural}. The failure of xNES for high-dimensional problems suggests that it is not advantageous to retain the full covariance matrix for optimization and that including the correlations between these parameters leads to bad performance. Thus, instead of considering the full covariance matrix for optimization of deep quantum circuits, a viable strategy could be using a block matrix correlating only some parameters, which we employ in section~\ref{sec:batch_opt}.
It is evident, that the rate of convergence of the optimization slows down with increasing number of parameters for both techniques which is expected as the search space becomes very large. 

\begin{figure}[htbp]
\centering
    \begin{tabular}{c c}
    \toprule
    10 layers & 20 layers\\
    \midrule
    \multicolumn{2}{c}{a) 5 qubits}\\
    \midrule
    \includegraphics[width=0.245\textwidth]{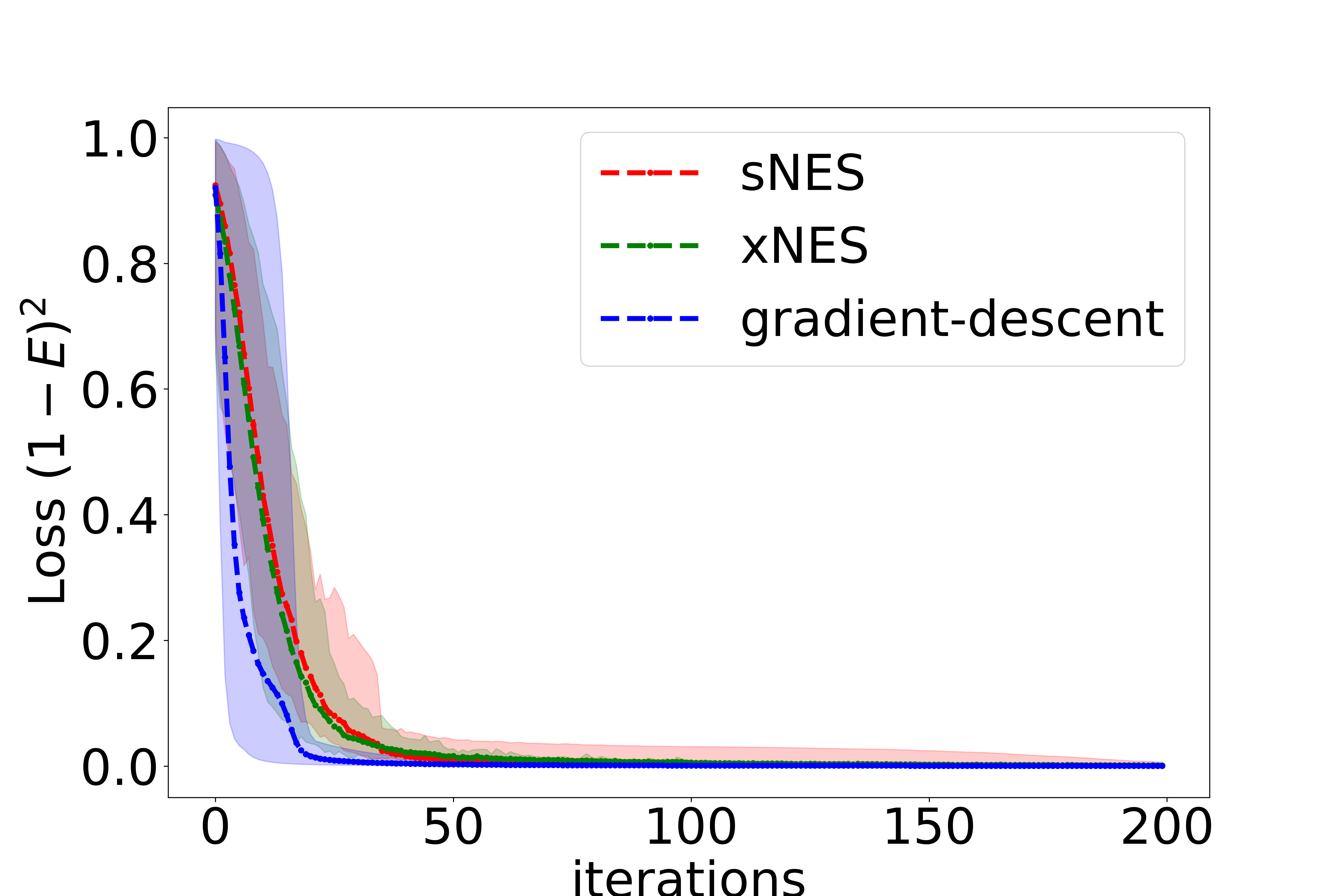} & 
     \includegraphics[width=0.245\textwidth]{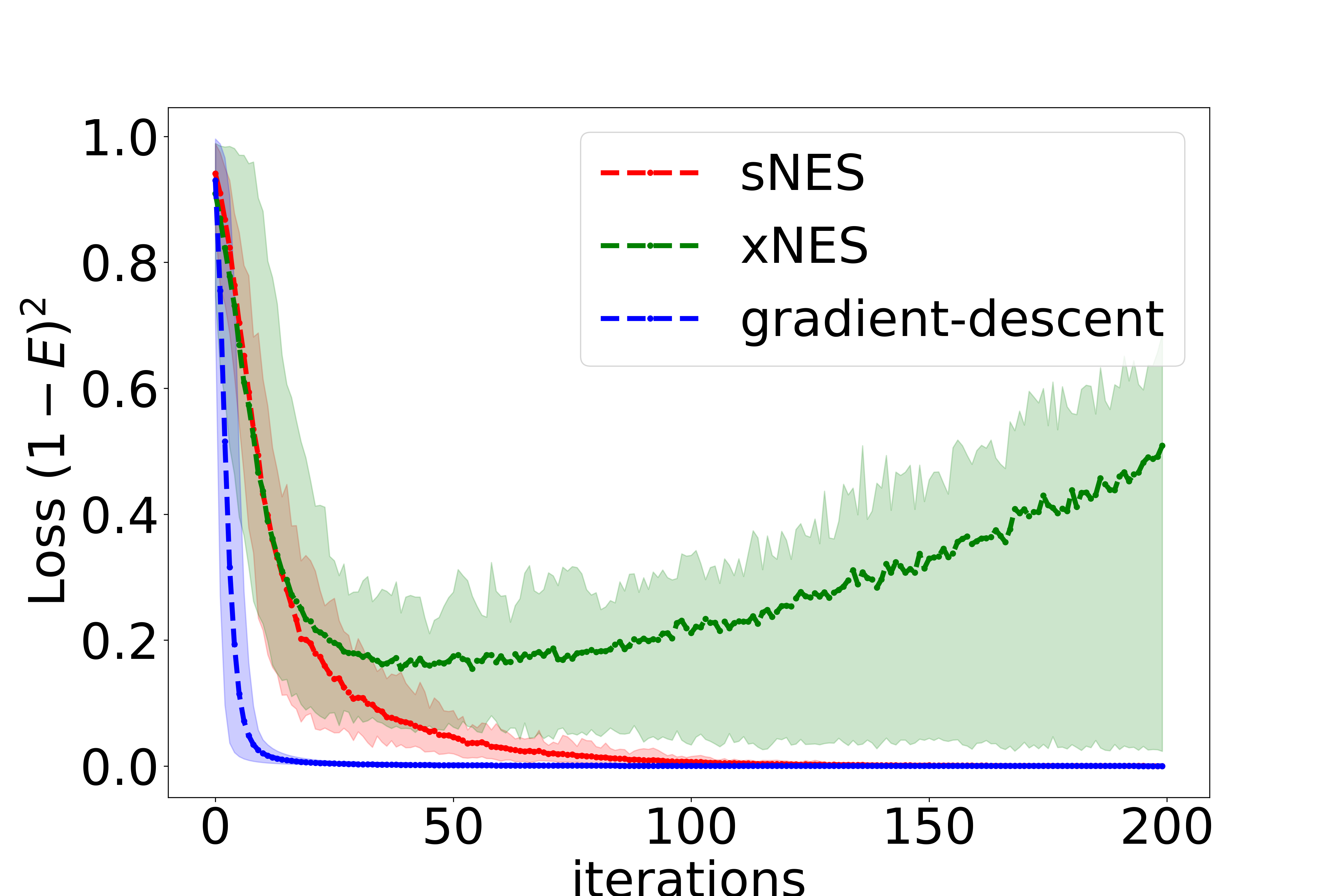}\\
     \midrule
    \multicolumn{2}{c}{b) 10 qubits }\\
    \midrule
    \includegraphics[width=0.245\textwidth]{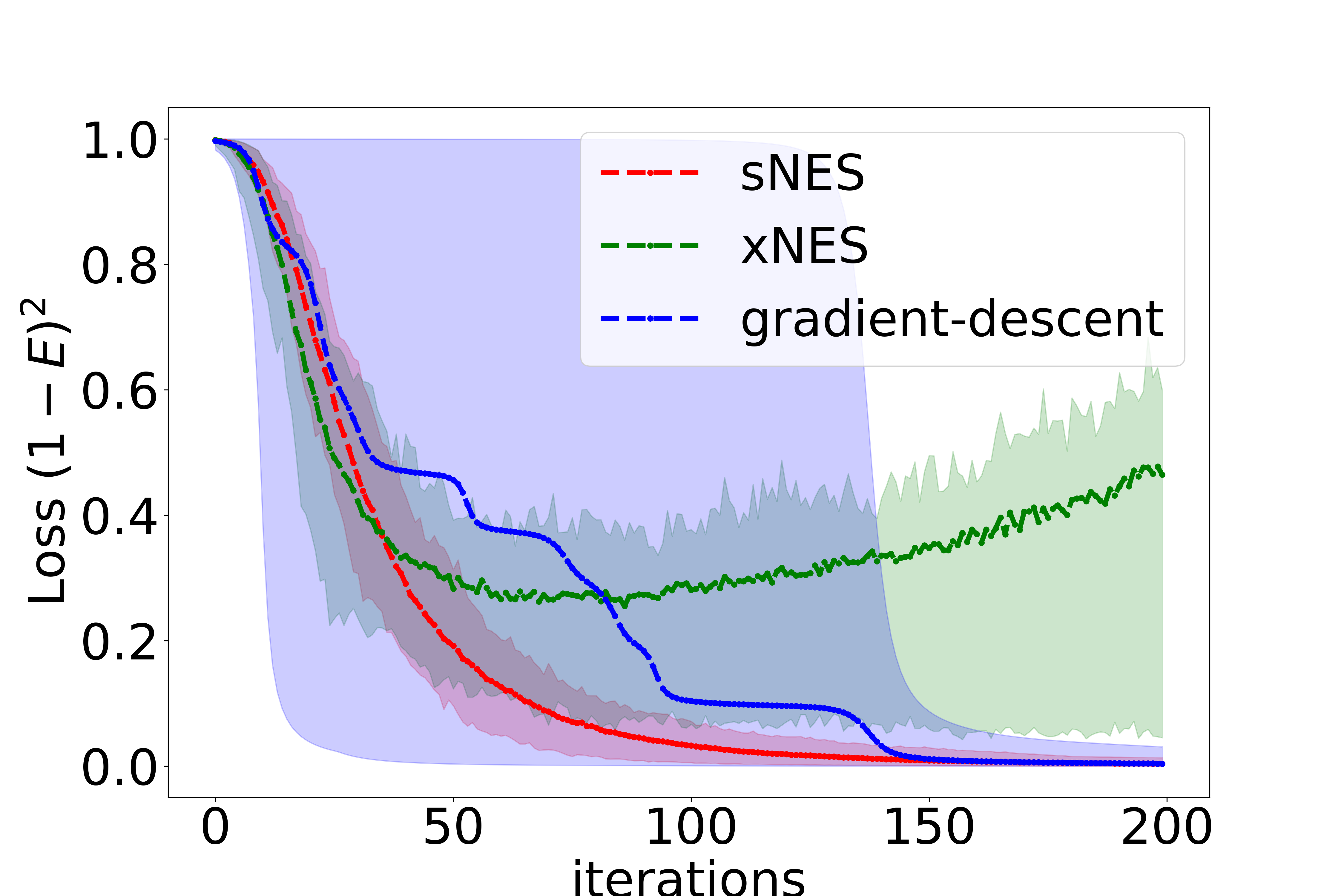} & 
     \includegraphics[width=0.245\textwidth]{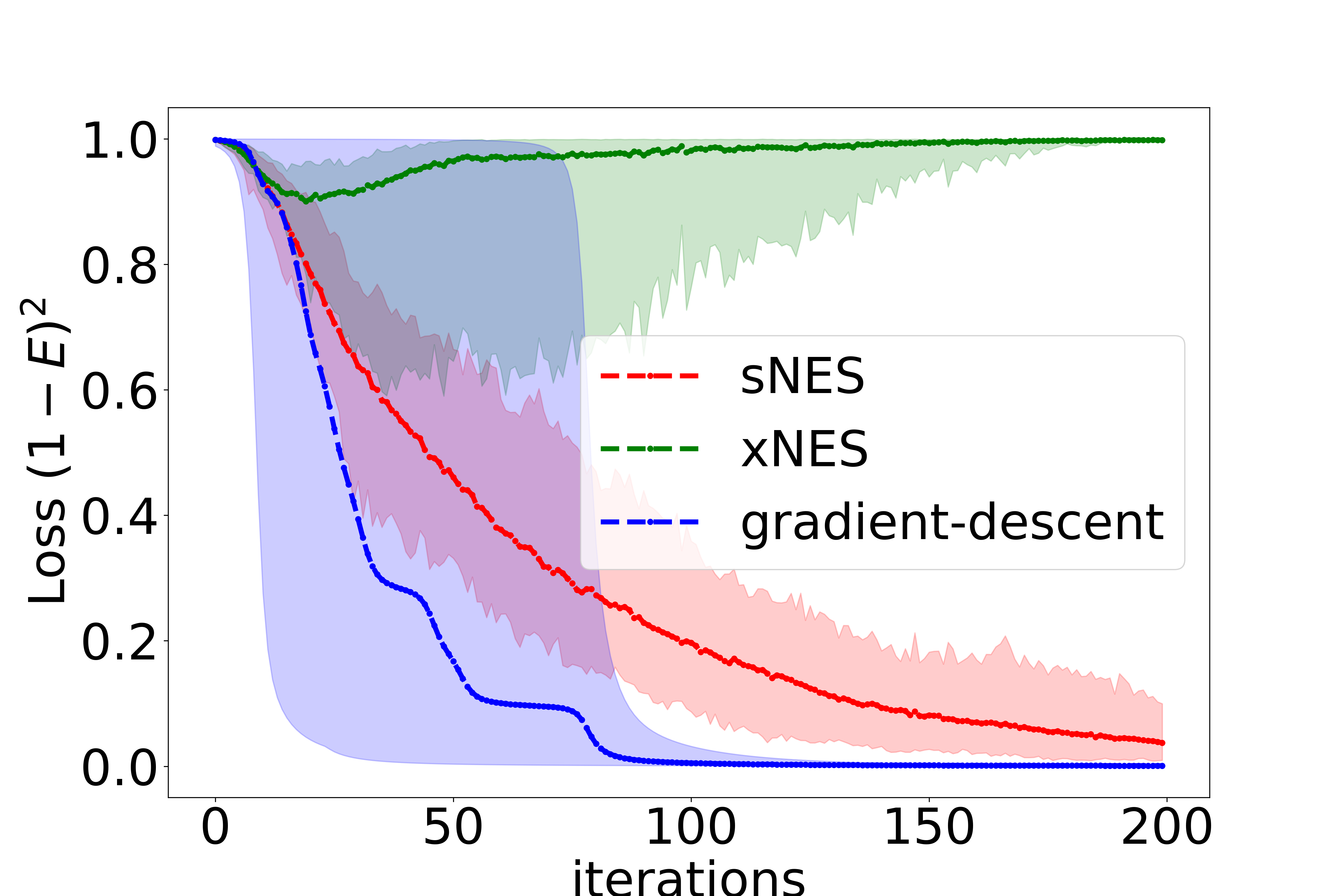}\\
     \midrule
    \end{tabular}
    \caption{Loss function with respect to the total number of iterations for the problem state preparation comparing optimization with different NES and gradient descent. The solid lines corresponds to the mean of the loss functions from 10 different runs, and the shadow represent the area between the best and worst values from the simulations.}
    \label{fig:com_w_grad_f}
\end{figure}

In Figure~\ref{fig:com_w_grad_f}, we compare the performance of NES with that of gradient descent for the optimization of RPQCs. The plots show that optimization with sNES performs equally well for all the cases in terms of finding the global minimum, but that xNES performance gets worse for circuits with more layers. An important point to make here is that, an update of the parameters using NES only requires $k$ circuit evaluations, while a standard gradient descent calculation with the parameter shift procedure typically needs twice the number of parameters ($2QL$) function evaluations.
Even though gradient descent reaches the global minimum in less iterations, it does so at a much higher cost in function evaluations. 
An interesting observation from the plots is that the initial rate of convergence with NES for circuits with 10 qubits or more is higher than that using gradient descent in many of the simulations, but similar on average. This is important in the context of the barren plateaus, where typically no suitable search direction can be found. 

It can be also seen from the plots that the NES work very similar for the different numerical simulations. However the variation in the loss values in simulations using gradient descent is very large indicating the need for a better strategies in the initial part of the optimization.
It opens the possibility for hybrid strategies, where the NES can be used to provide the inital direction for gradient descent for optimizing large RPQCs.

\subsection{VQE for the H$_2$O molecule}
\begin{figure}[htbp]
\centering
     \includegraphics[width=0.475\textwidth]{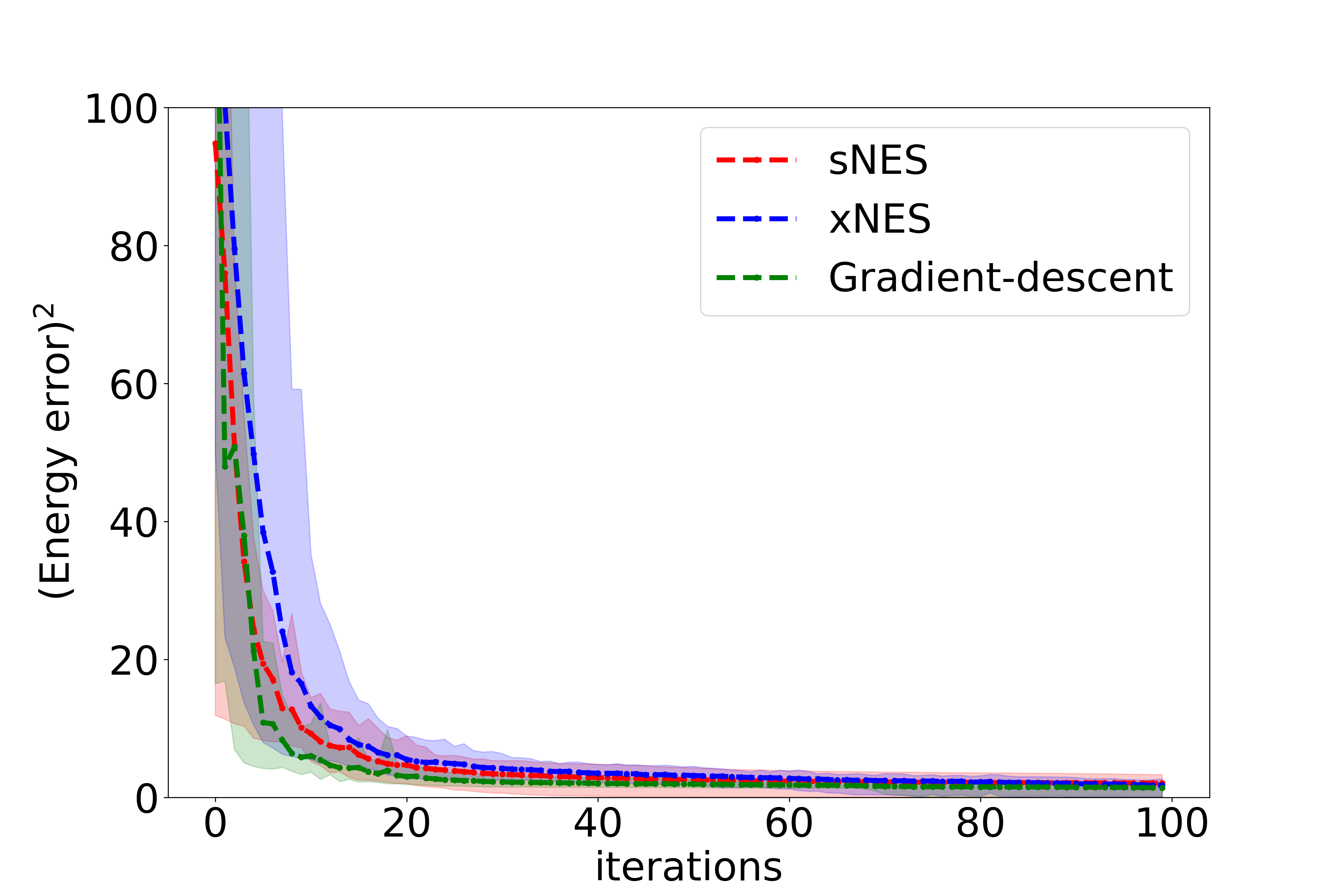}
    \caption{Convergence to the ground state energy for H$_2$O molecule using VQE with randomly initialized UCC ansatz. The solid lines corresponds to the mean of the loss functions from 10 different runs, and the shadow represent the area between the best and worst values from the simulations.}
    \label{fig:vqe_water}
\end{figure}

In the context of chemistry, PQCs are commonly used to represent the wave function of molecules.
Although the circuits here typically use structured physically motivated ansatze, the number of parameters becomes large as the system size grows.
The motivation behind choosing this optimization problem is to investigate the applicability of NES to problems with physically inspired ansatze, and compare their performance with gradient-based methods for such problems.
In this simulation we estimate the ground state energy of a water molecule at equilibrium geometry using the STO-3G basis (14-qubit) with the UCCSD ansatz (34 parameters).
The parameters of the circuit are intitialized randomly, which in certain cases lead to convergence issues due to presence of barren plateaus or getting stuck in a local minina as observed in~\cite{cervera2020meta}.
We plot the results from the numerical experiments in Figure~\ref{fig:vqe_water}, and it can be seen from the plots that we can use both sNES and xNES to optimize the parameters for estimating the ground state energy of water molecule using VQE. We also carry out the optimisation with a gradient-based optimizer, and plot the results in Figure~\ref{fig:vqe_water}. 
It can be seen from the plots that the performances of sNES, xNES and gradient descent are pretty similar.
However, the xNES performs equally well for this example as sNES, even though the size of the water molecule is beyond the limit for which we previously found that xNES loses performance compared to sNES. 
This suggests that in this case keeping the covariance matrix gives an advantage and that the parameters for this problem are actually correlated.
The degree of randomness in the circuit topology can be a factor in selecting the optimal NES flavor. 

\subsection{Batch optimization}\label{sec:batch_opt}

As seen in Figure~\ref{fig:full_s_nes} the rate of convergence decreases with increasing number of parameters.
This leads to long optimization runs for growing circuits.
At the same time the overhead of the full NES strategy grows.
To alleviate these effects, we investigate the use of batch optimization strategy for optimizing RPQCs of larger depth, as investigated in another recent study~\cite{skolik2020layerwise} where the authors investigate the use of layerwise learning techniques for optimisation of Quantum Neural Networks. 
In batch optimization, only a limited number (the batch size) of parameters are adjusted every iteration to lower the cost.
We carry out various simulations to analyze the effect of the size of the batches, number of walkers and different partitioning strategies on the convergence of the optimization.

\begin{figure}[htbp]
\centering
    \begin{tabular}{c c}
    \toprule
    sNES & xNES\\
    \midrule
    \multicolumn{2}{c}{a) Batch size of 5}\\
    \midrule
    \includegraphics[width=0.245\textwidth]{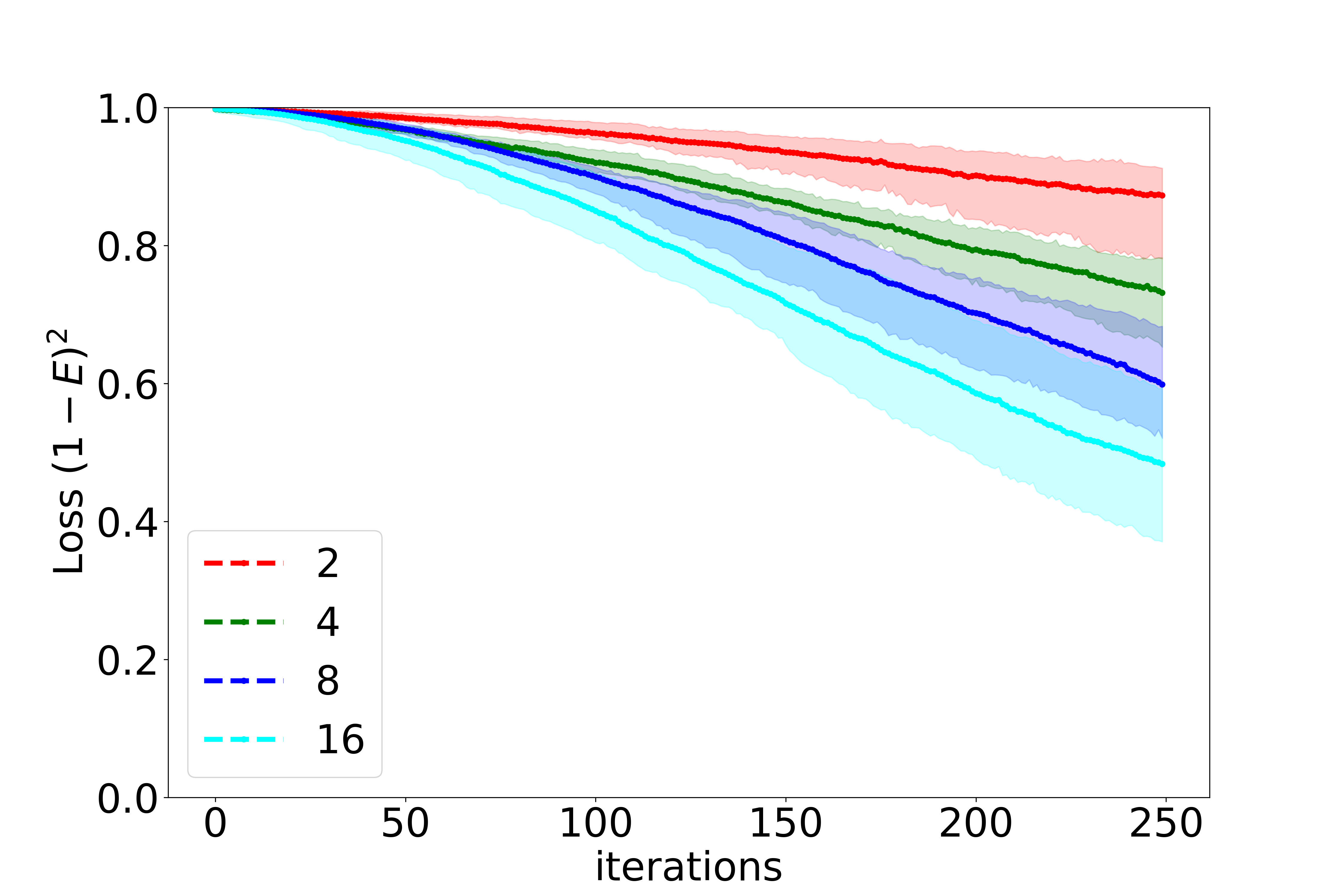} & 
     \includegraphics[width=0.245\textwidth]{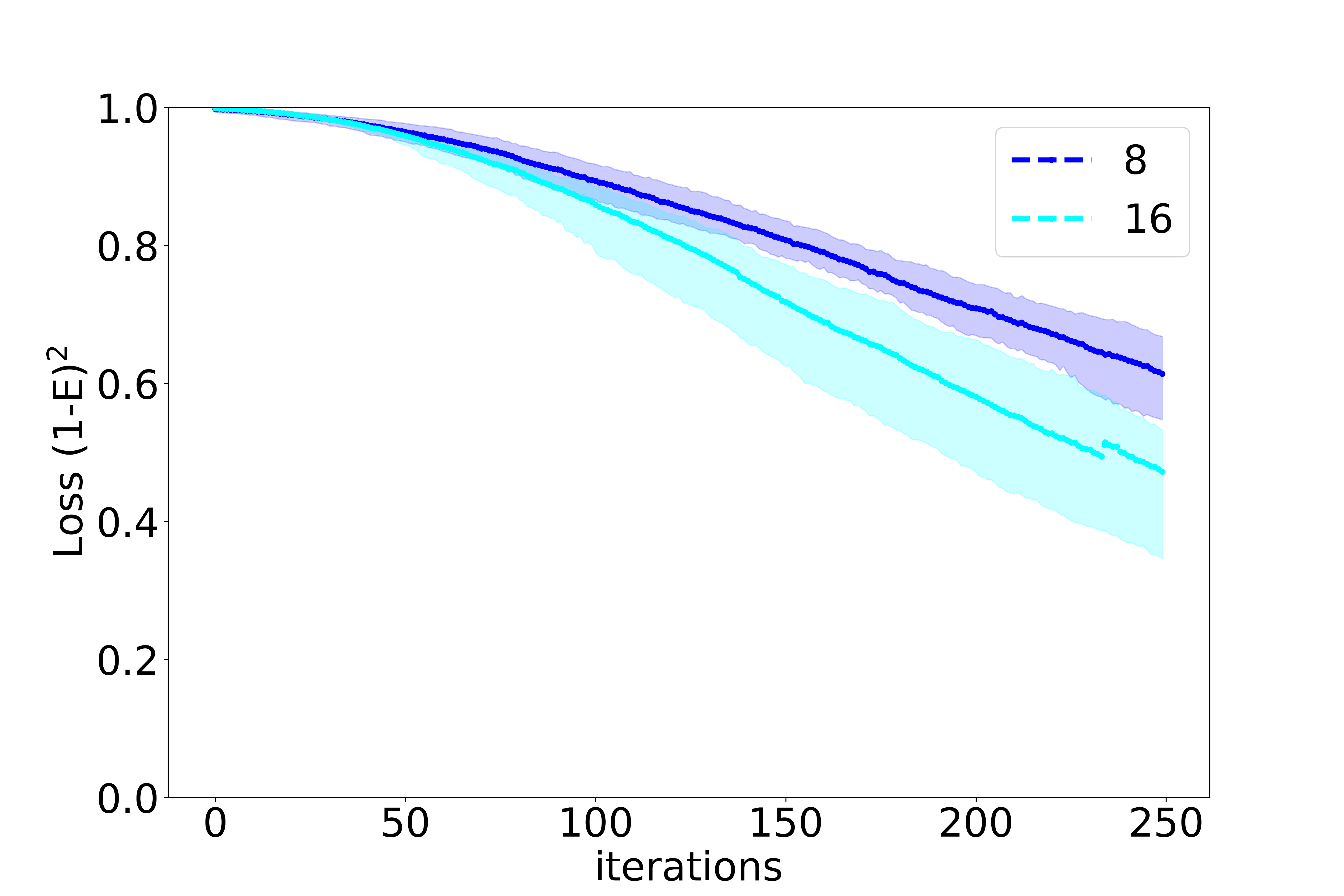}\\
     \midrule
    \multicolumn{2}{c}{b) Batch size of 25}\\
    \midrule
    \includegraphics[width=0.245\textwidth]{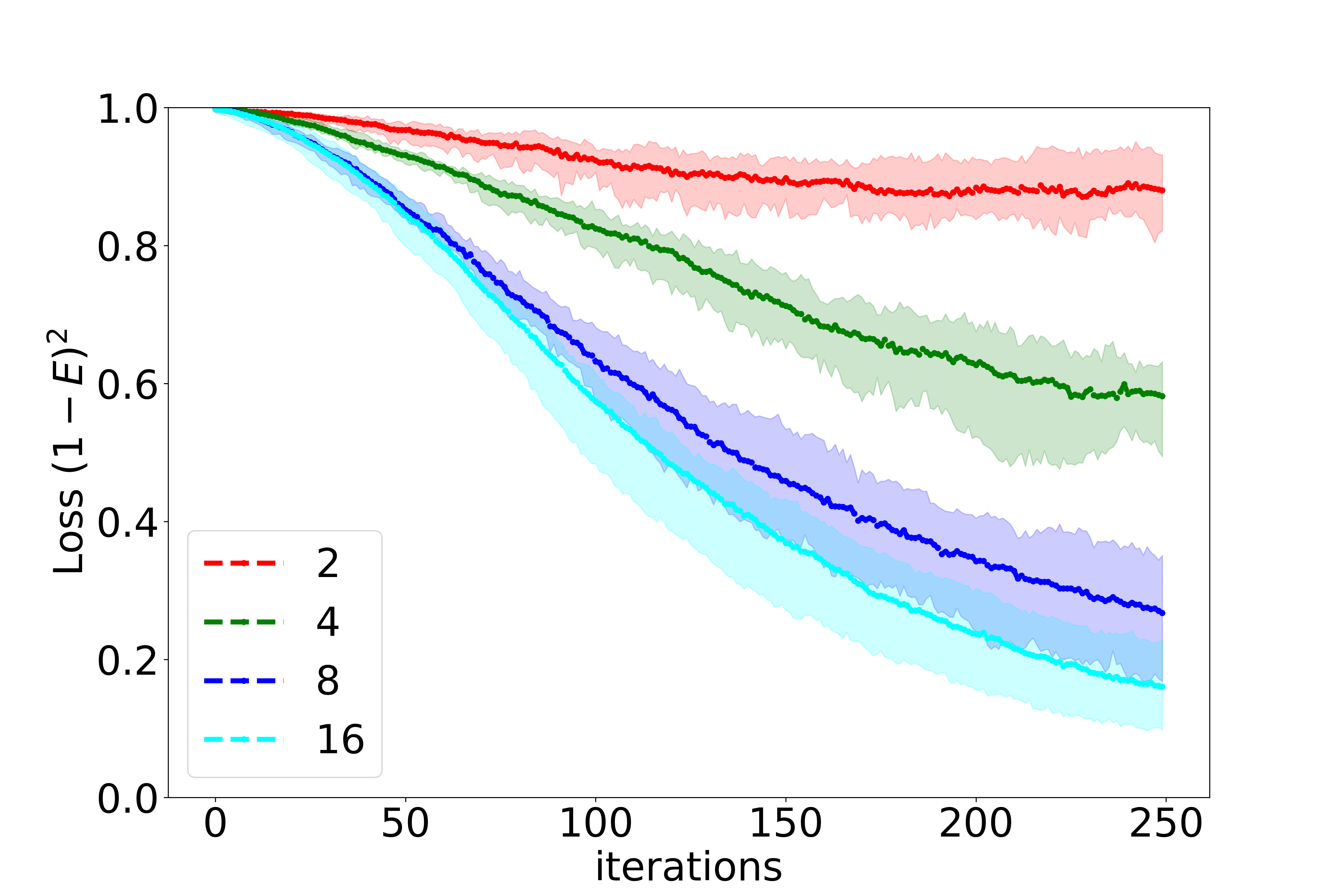} & 
     \includegraphics[width=0.245\textwidth]{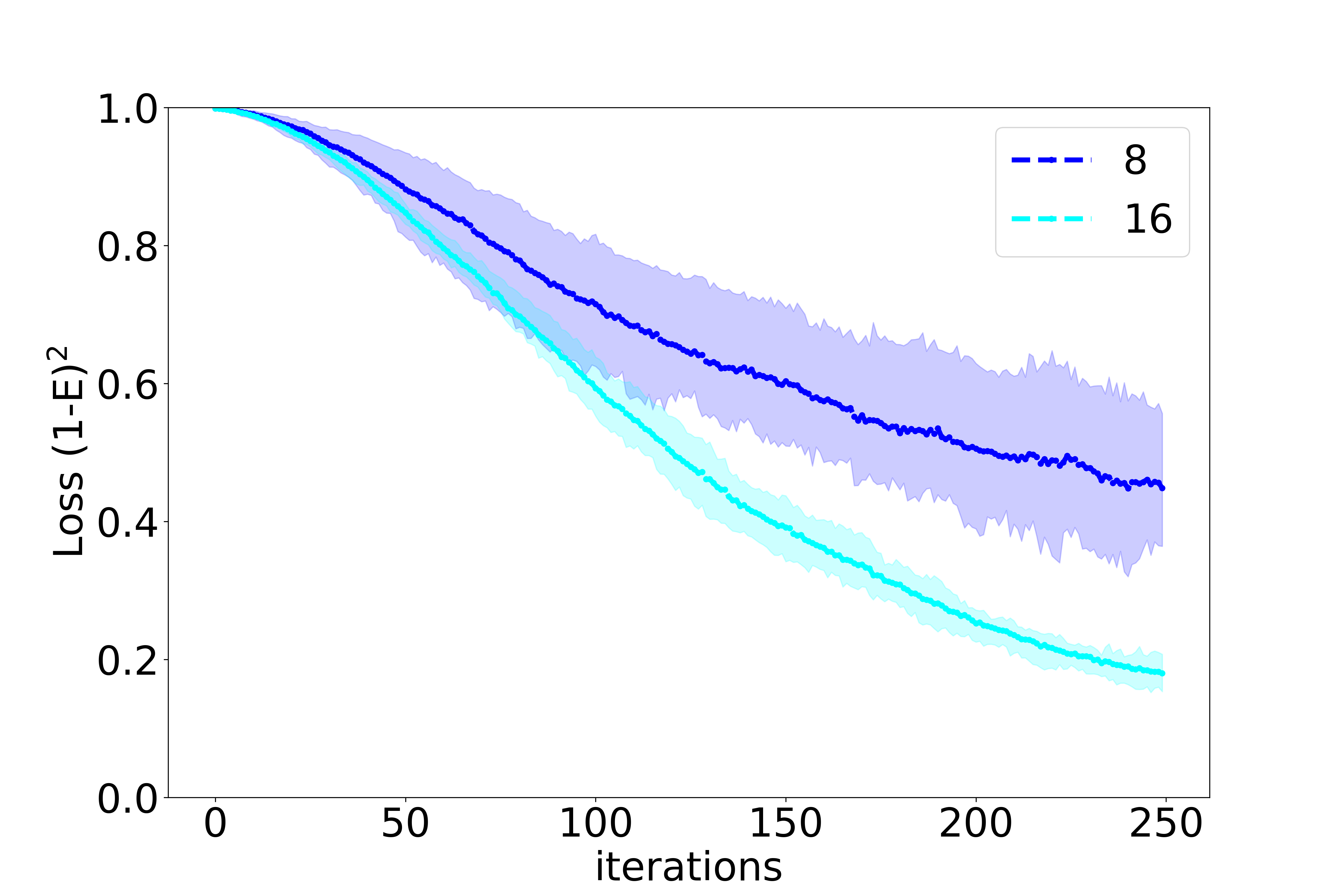}\\
     \midrule
    \multicolumn{2}{c}{c) Batch size of 50}\\
    \midrule
    \includegraphics[width=0.245\textwidth]{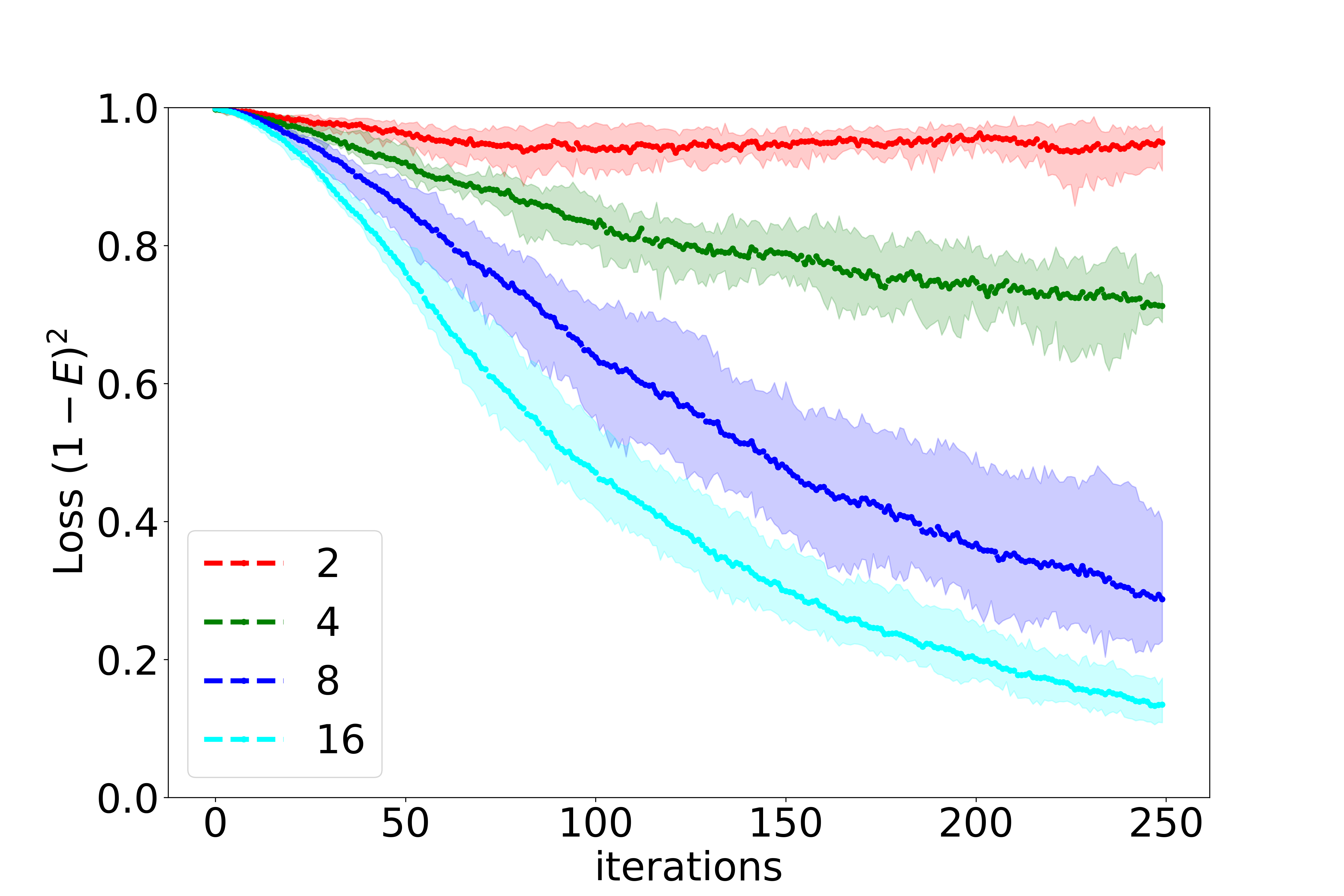} & 
     \includegraphics[width=0.245\textwidth]{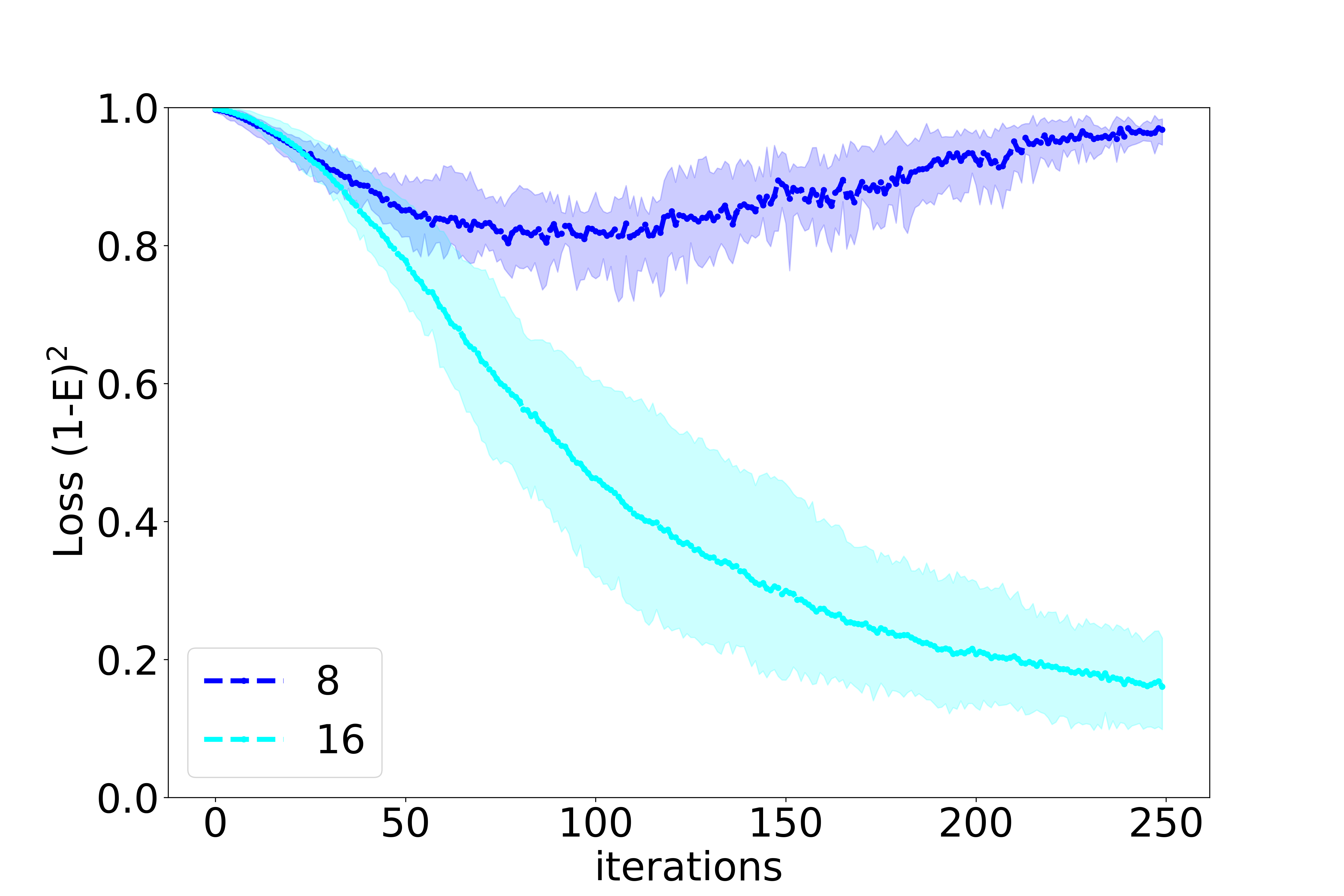}\\
     \midrule
  
    \end{tabular}
    \caption{Loss function with respect to the total number of iterations of batch optimization. The solid lines corresponds to the mean of the loss functions from 6 different runs, and the shadow represent the area between the best and worst values from the simulations. The various colors indicate different number of walkers used for optimization.}
    \label{fig:diff_bs_nw}
\end{figure}

We first investigate the effect of the size of batches and the number of walkers on the optimization by carrying out different numerical simulations with varying batch size. 
We use RPQCs with 10 qubits and 50 layers for the simulations and the batches of parameters are selected randomly but fixed over the entire calculation. 
The results in Figure~\ref{fig:diff_bs_nw} show clearly that a minimum batch size is required.
The xNES needs more walkers to give meaningful results than the sNES.
The curves show both a convergence in the batch size and the number of walkers at around 16 walkers and a batch size of 50.
We will use these values in the following calculations, it is expected that these values have some problem dependence and need to be tuned accordingly.

\begin{figure}[htbp]
\centering
    \includegraphics[width=0.475\textwidth]{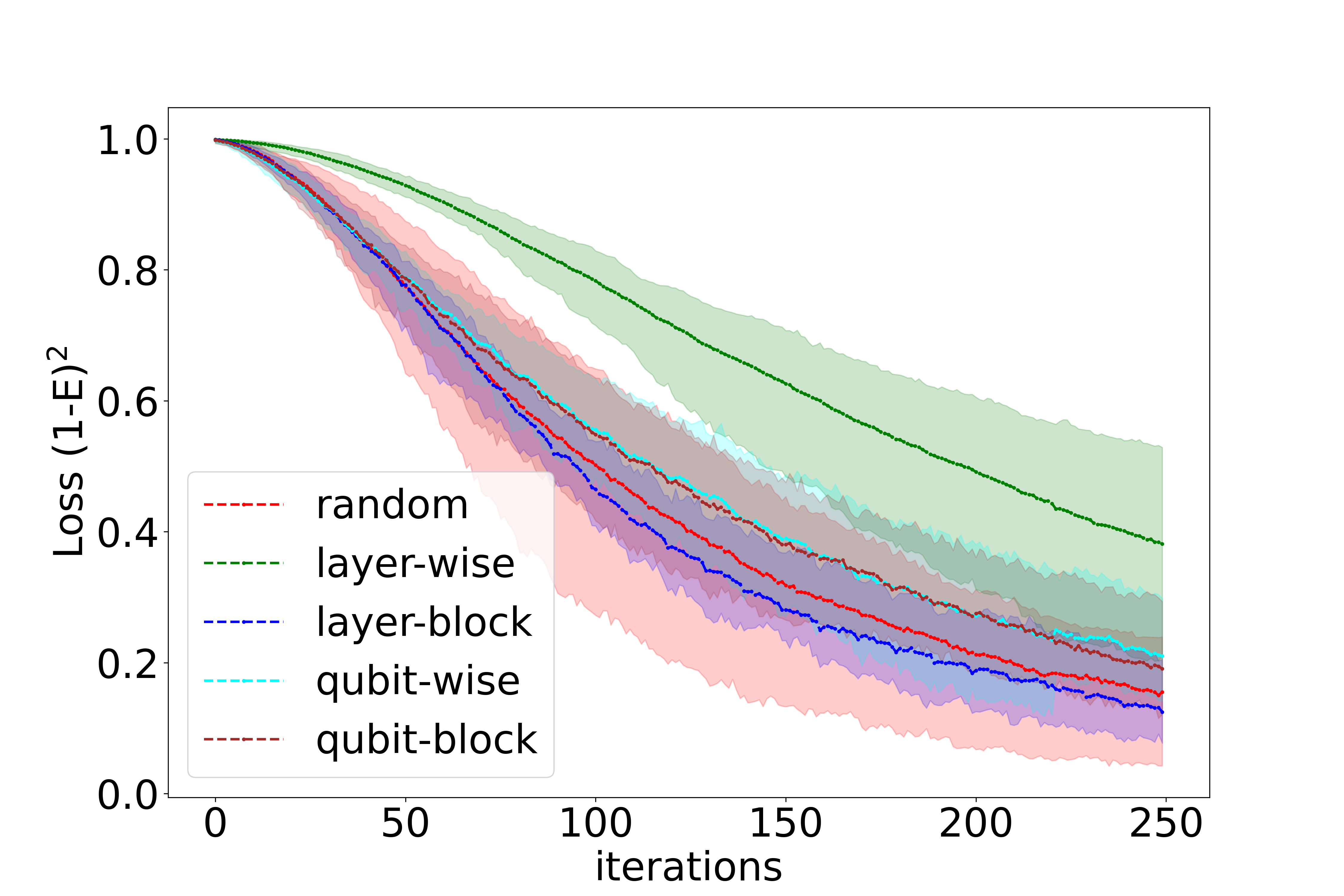} 
    \caption{Loss function with respect to the total number of iterations using different strategies for batch optimization using sNES. The solid lines corresponds to the mean of the loss functions from 6 different runs, and the shadow represent the area between the best and worst values from the simulations.}
    \label{fig:diff_batches_s}
\end{figure}

We next carry out simulations to investigate the effect of different strategies for generating the batches. The different strategies we investigate are dividing the circuit parameters randomly, layer-wise, qubit-wise, layer-block-wise, and qubit-block-wise. We plot the results from simulations of 10 qubit 50 layered RPQCs in Figure~\ref{fig:diff_batches_s}. It can be seen from the plot that the curves are identical for all the strategies, with the layer-wise batch optimization strategy a bit deviant because of the different batch size. We also carried out different simulation with RPQCs of larger depths using the different partition strategies and found that depending on the problem one might get faster convergence using one of the strategies, thus we recommend investigating the use of different strategies when dealing with larger RPQCs. Other techniques such as the parameter efficient circuit training techniques~\cite{sim2020adaptive} can be also used together with NES to further improve the optimization procedure.

\begin{figure}[htbp]
    \includegraphics[width=0.475\textwidth]{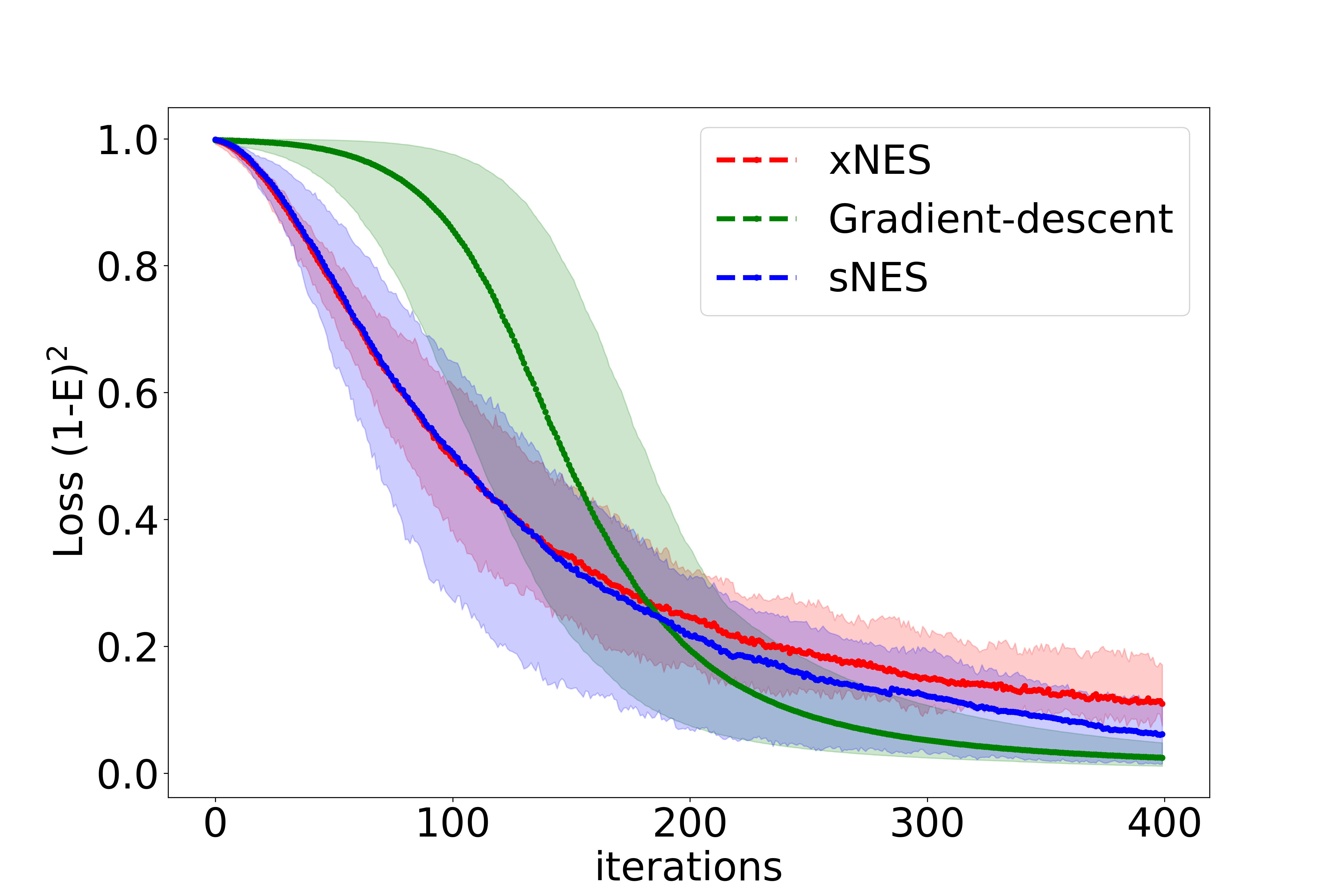} 
    \caption{Loss function with respect to the total number of iterations of batch optimization comparing the performance of the NES with gradient descent. The solid lines corresponds to the mean of the loss functions from 10 different runs, and the shadow represent the area between the best and worst values from the simulations.}
    \label{fig:b_grad_comp}
\end{figure}

We also analyzed the performance of NES with gradient descent with the batch optimization strategy. We use RPQCs with 10 qubits and 50 layers, and use batch size of 50 parameters and 16 walkers for the simulations. The results from the simulation is plotted in Figure~\ref{fig:b_grad_comp}. The optimization with NES performs better than gradient descent initially, as seen previously in Figure~\ref{fig:com_w_grad_f}. 

\begin{figure}[htbp]
\centering
    \begin{tabular}{c c}
    \toprule
    sNES & xNES\\
    \midrule
    \multicolumn{2}{c}{a) 10 qubits RPQCs}\\
    \midrule
    \includegraphics[width=0.245\textwidth]{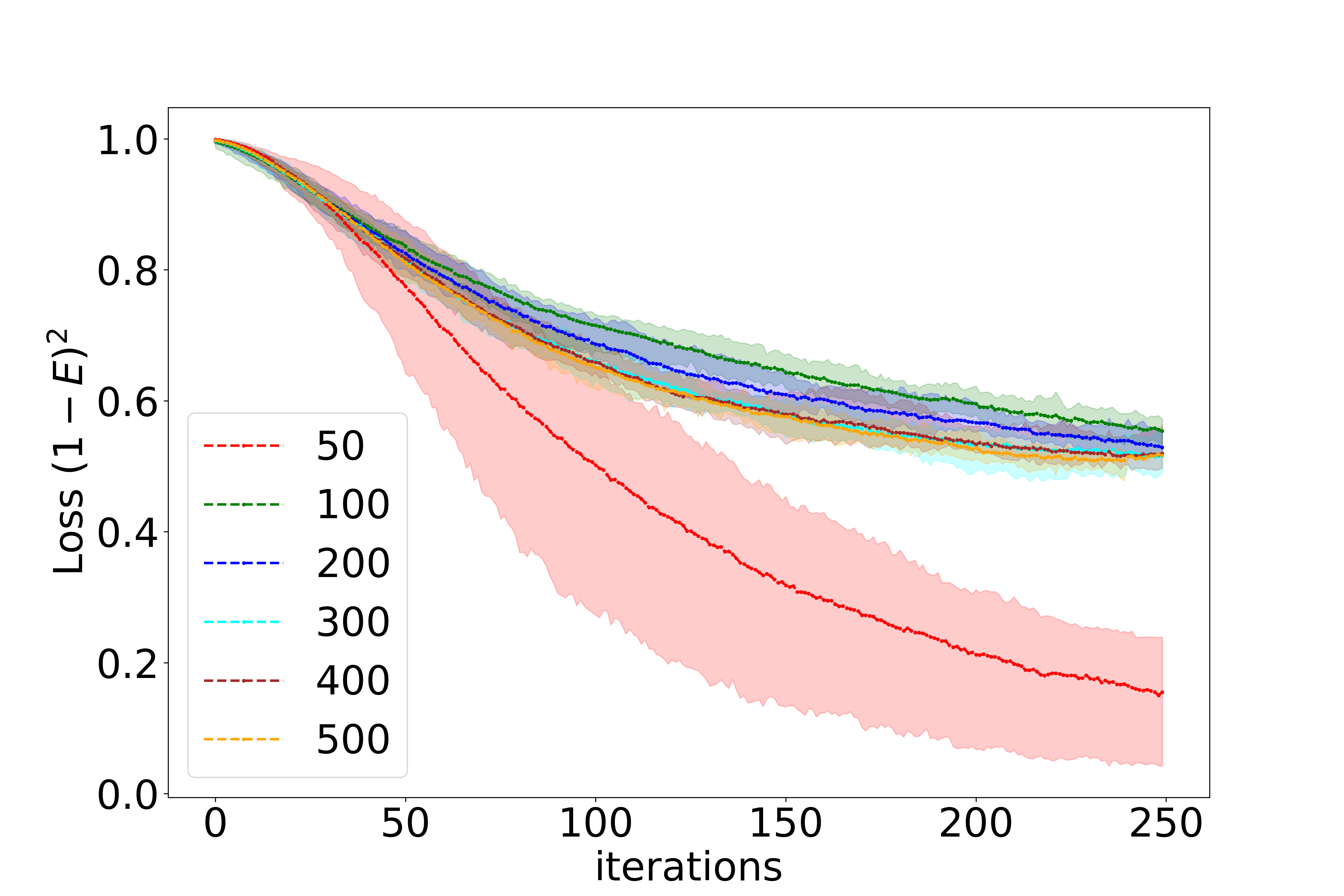} &
     \includegraphics[width=0.245\textwidth]{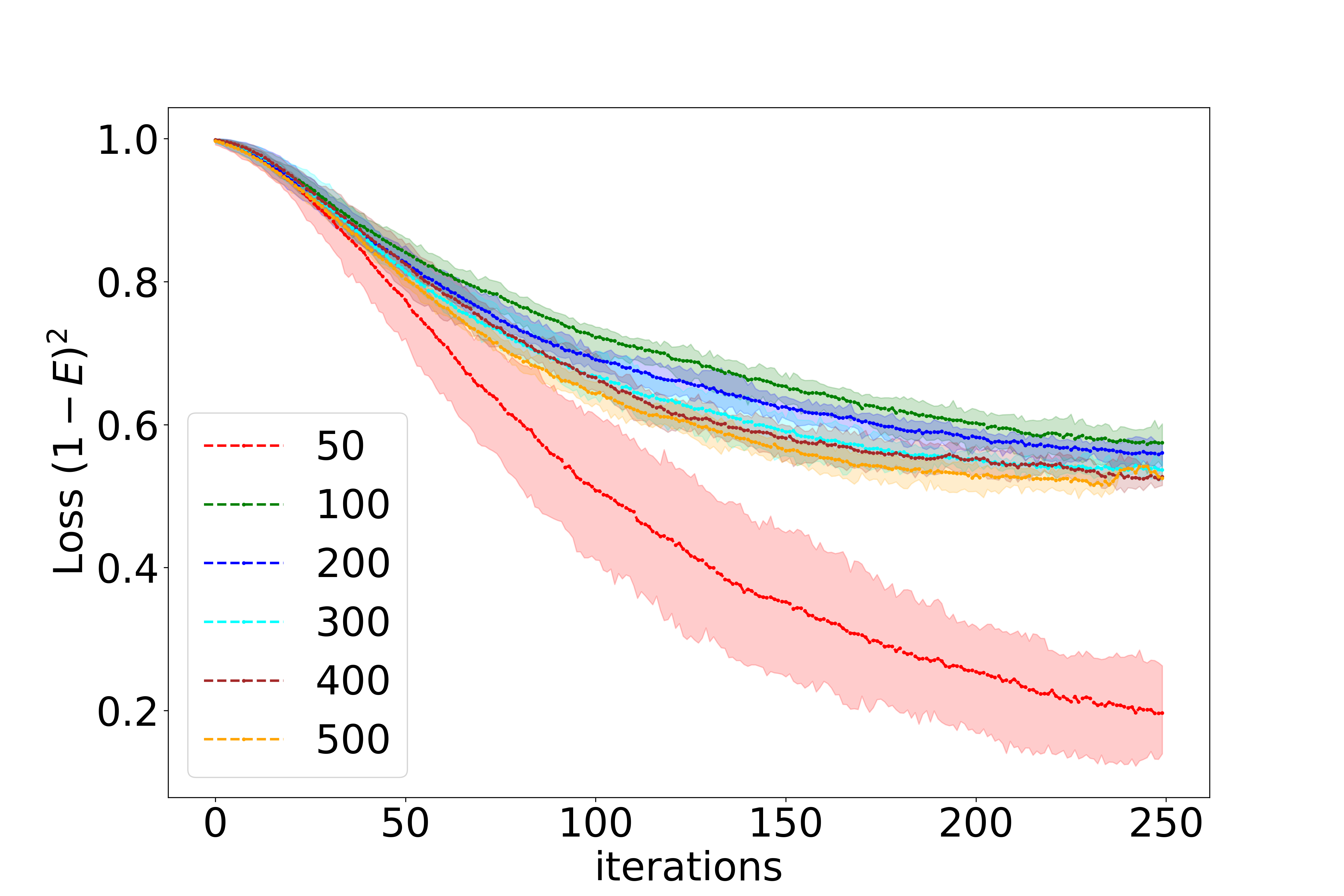}\\
     \midrule
    \multicolumn{2}{c}{b) 15 qubits RPQCs}\\
    \midrule
    \includegraphics[width=0.245\textwidth]{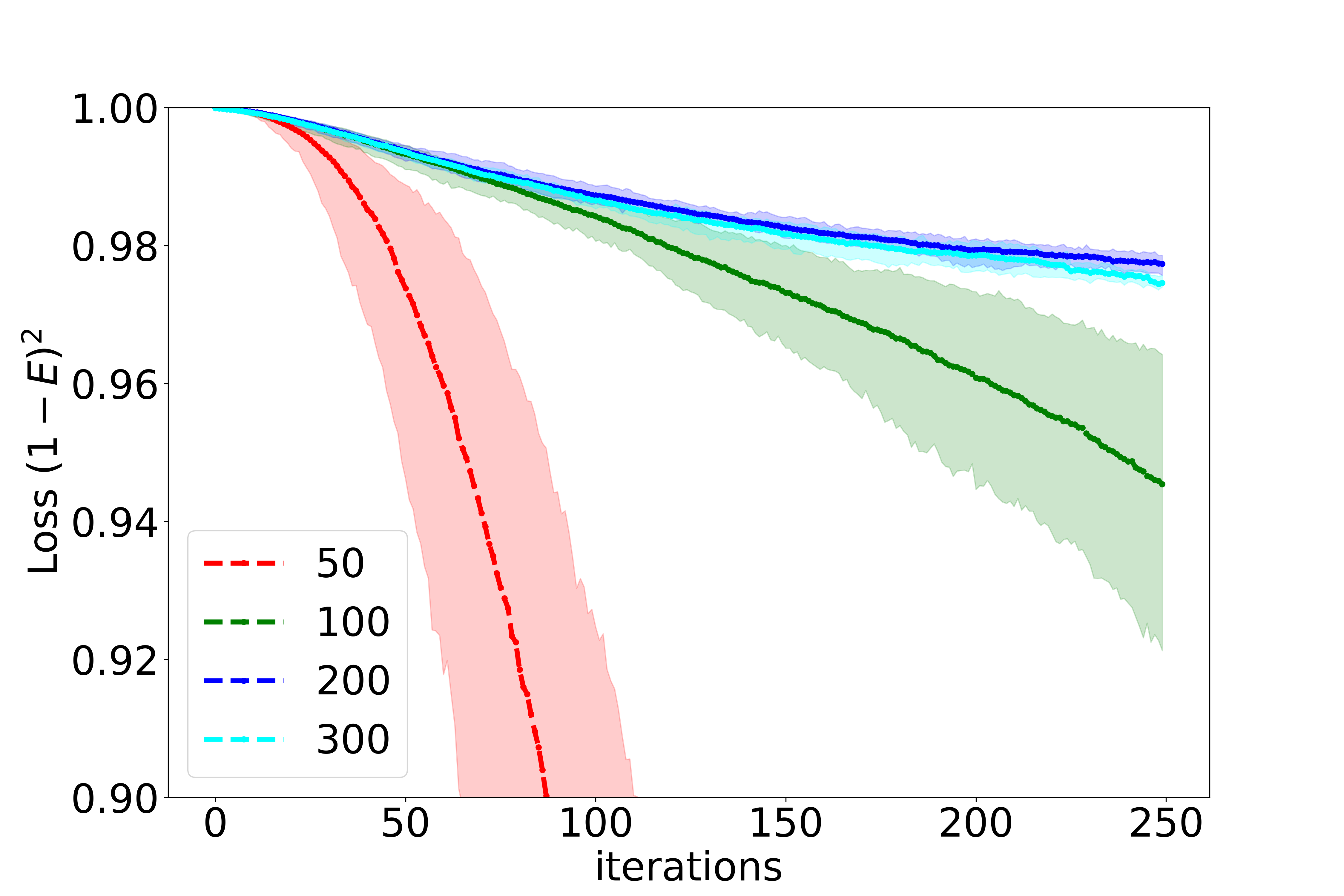} & 
     \includegraphics[width=0.245\textwidth]{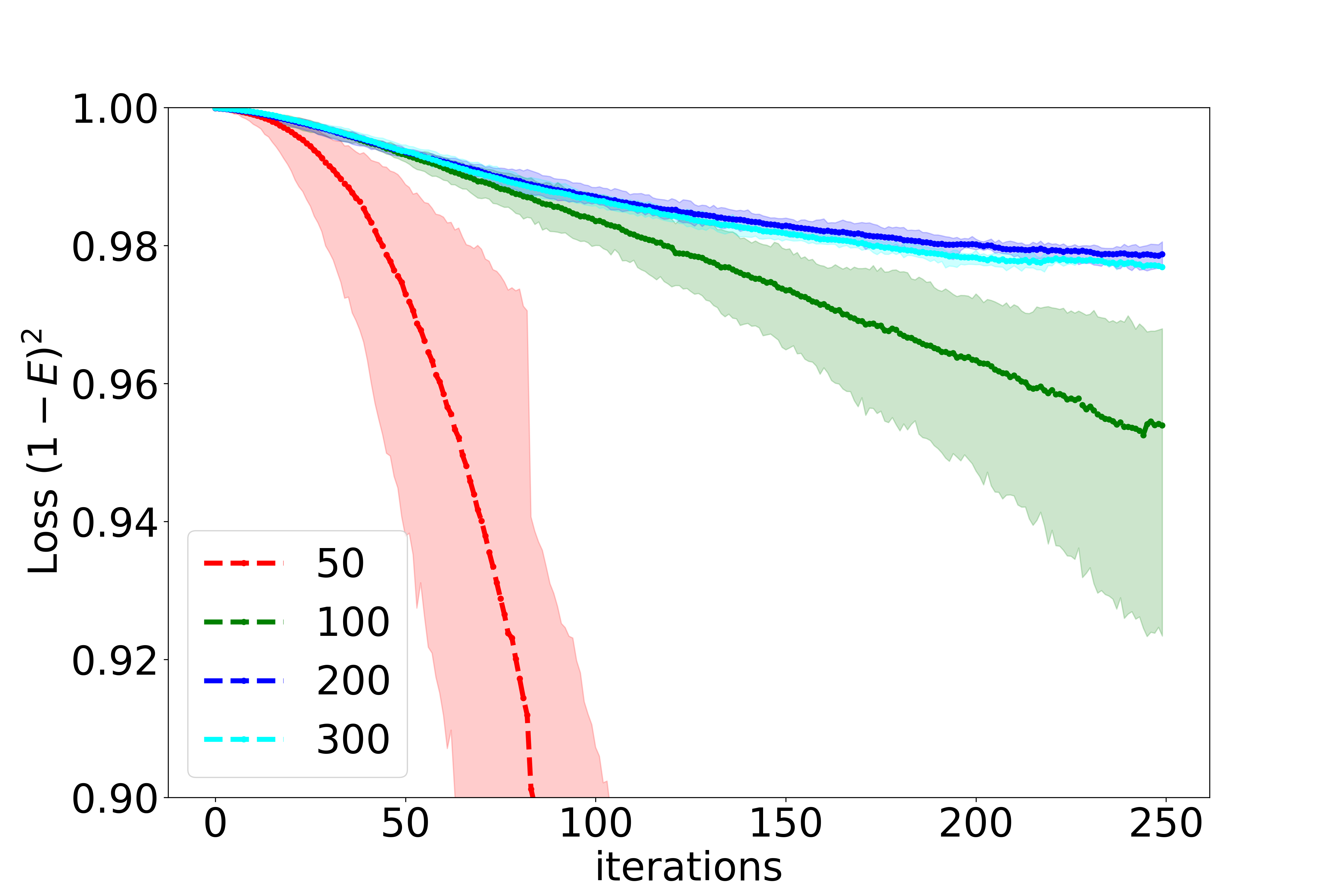}\\
     \midrule
  
    \end{tabular}
    \caption{Loss function with respect to the total number of iterations of batch optimization. The solid lines corresponds to the mean of the loss functions from 10 different runs, and the shadow represent the area between the best and worst values from the simulations. The various colors indicate different number of layers in the quantum circuit.}
    \label{fig:s_x_nes}
\end{figure}

Finally we ran simulations to carry out the optimization of RPQCs with varying number of qubits and layers using both xNES and sNES by randomly dividing the parameters in batches of size 50 and using 16 walkers. We show the results for 10-qubit and 15-qubits circuits with different number of layers in Figure~\ref{fig:s_x_nes}. The plots show that we can perform optimization of significantly deep quantum circuits. As the circuits depth increases the rate of convergence decreases, and one might require a high number of circuit runs for the final result. Faster convergence of deep RPQCs might be achieved by selecting specific values of different parameters such as the initial learning rates, initial values of covariance matrix, number of walkers and the batch size. We leave such hyperparameter tuning for a future study.

\subsection{Hybrid strategy}
\begin{figure}[htbp]
\centering
    \begin{tabular}{c}
    \toprule
    a) 10 qubits 50 layers RPQC \\
    \midrule
    \includegraphics[width=0.475\textwidth]{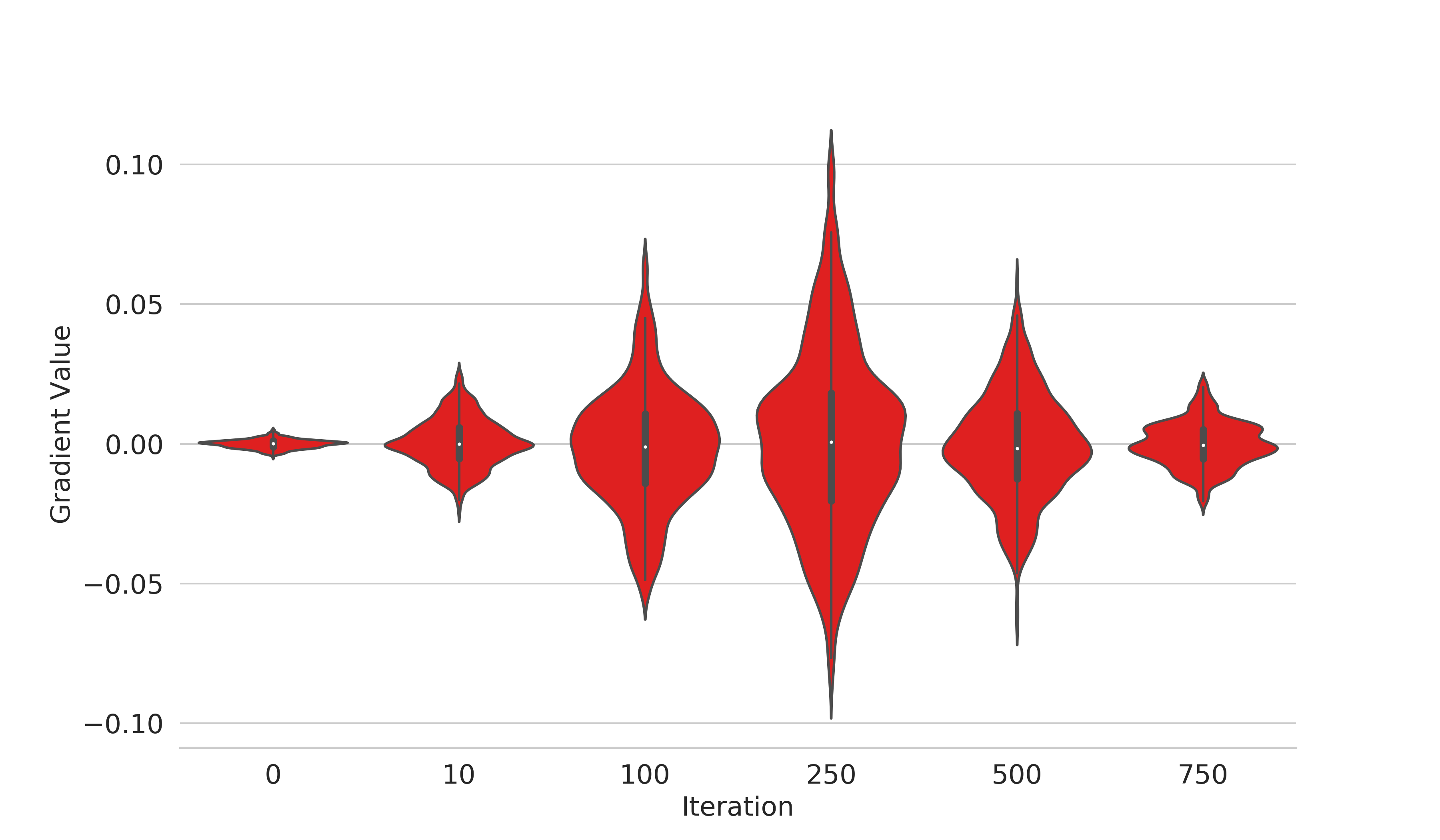} \\
    \midrule
    b) 15 qubits 50 layers RPQC \\
    \midrule
    \includegraphics[width=0.475\textwidth]{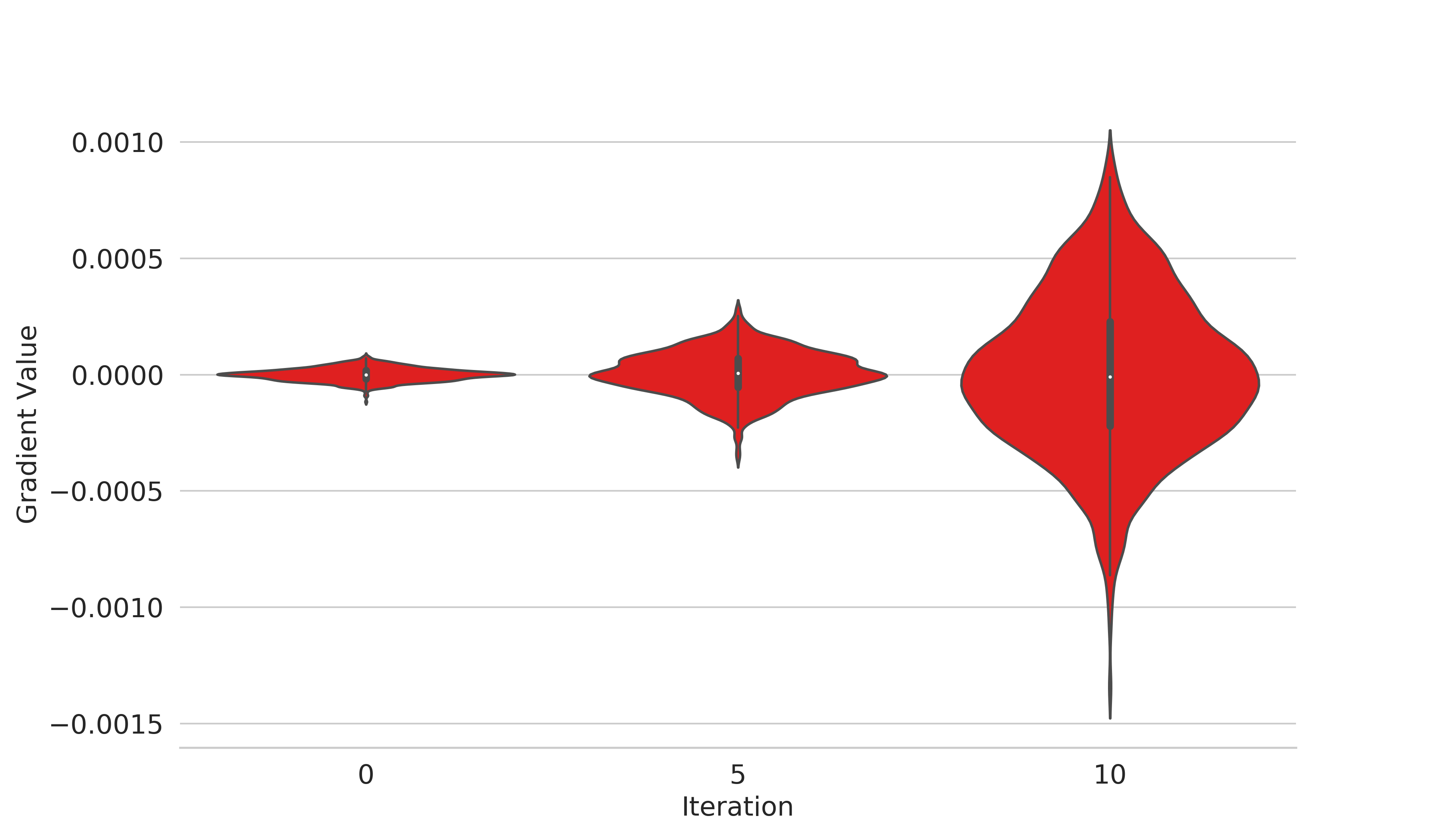} \\
    \midrule
    c) 18 qubits 10 layers RPQC \\
    \midrule
    \includegraphics[width=0.475\textwidth]{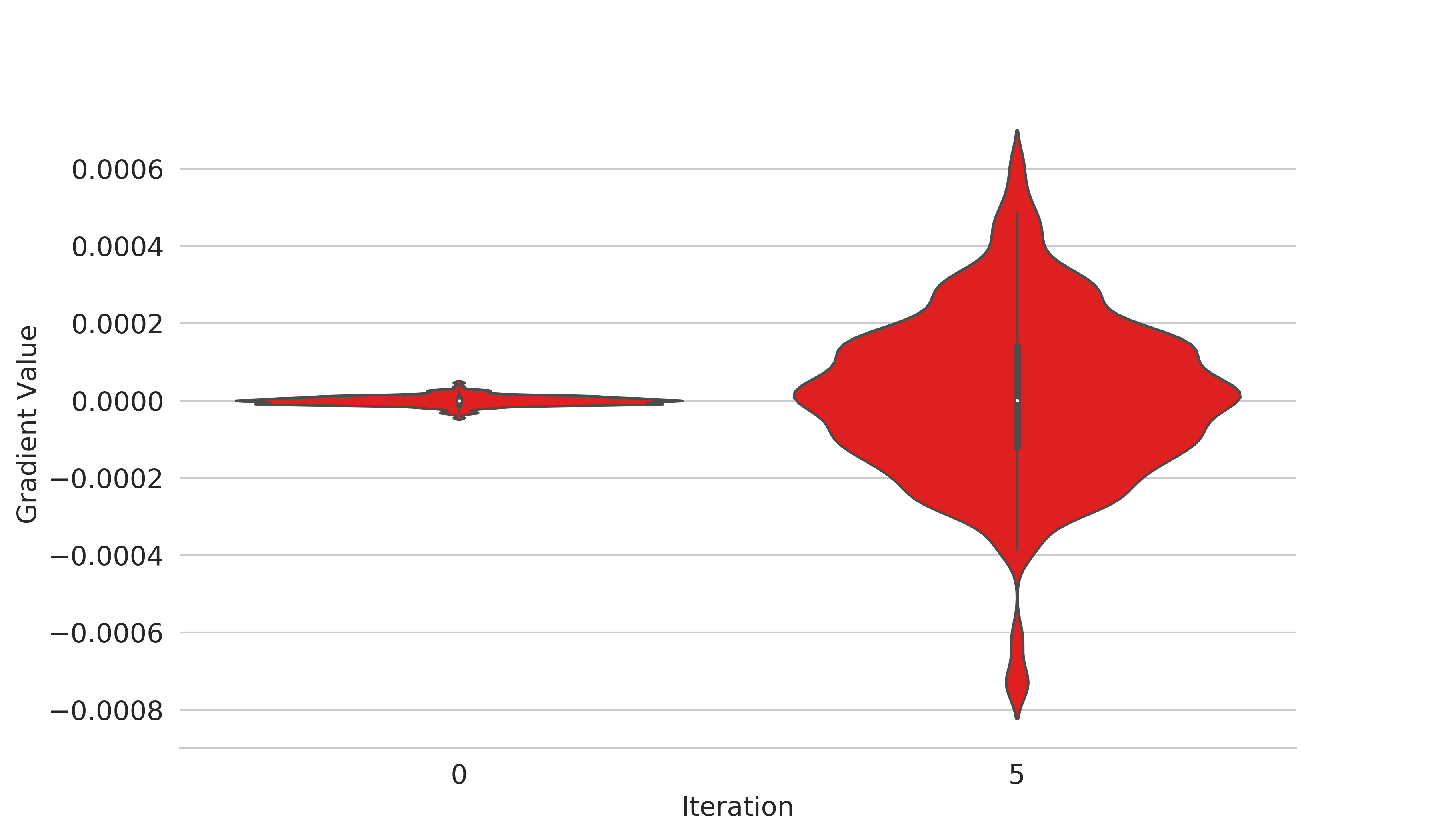} \\
    \midrule
    
    \end{tabular}
    \caption{A violin plot illustrating the distribution of the analytical gradient with respect to all the parameters of the quantum circuit at different iterations in the optimization using NES.}
    \label{fig:hybrid}
\end{figure}

As shown above, the convergence with NES is sometimes slower than gradient-based methods in regions with non-vanishing gradients. 
In this section we explore a hybrid strategy that uses separable natural evolution strategies (sNES) in areas with vanishing gradients and regular gradient descent when a search direction has been found. 
For this, we simulated circuits with 10 qubits 50 layers, 15 qubits 50 layers, and 18 qubits 10 layers RPQCs with 500, 750 and 180 parameters respectively.
We provide numerical evidence of the viability of this strategy by tracking the value of the analytical gradient with respect to all the parameters at different iterations in the optimization with NES, and plot the resulting distribution of the analytical gradients using violin plots in Figure~\ref{fig:hybrid}. 

We tracked the analytical gradient distribution for the full optimization run for the 10 qubit 50 layers RPQC plotted in Figure~\ref{fig:hybrid}a, and as expected, the distributions follow the pattern where it spreads initially to larger values and then narrows down about zero near the end of the optimization. This indicates that if one can increase the spread of the gradients at the beginning of the optimization one might be able to use gradient-based methods to complete the process. Thus, we generate the distribution of the analytical gradients after every 5 iteration of optimization by sNES, and plot the distributions for 15 and 18 qubits RPQCs in Figure~\ref{fig:hybrid}b \& \ref{fig:hybrid}c. The distribution of the analytical gradient in Figure~\ref{fig:hybrid}b \& \ref{fig:hybrid}c start to spread to higher values already after 5 iterations of optimization with sNES, which indicates that we can then use gradient-based methods in these cases for the full optimization. The choice of the number of optimization step will depend more on the problem, and one can use sNES until the magnitudes of the analytical gradients are large enough. These empirical results indicate that using a surrogate gradient as the one in sNES, one might be able to navigate through areas of vanishing gradients and then use gradient-based methods for finding the optimal solution.

\section{Conclusion and Outlook}\label{sec:conclusion}
We presented a strategy of representing parameters of quantum circuits as multinormal distributions for parameterized quantum circuits and use natural evolution strategies for optimizing randomly initialized PQCs. We have shown that the search direction used in NES can be amplified for randomly initialized parameterized quantum circuits, and that with appropriate number of walkers one can use them for optimization in all cases. We numerically analyze the performance of NES for parameterized quantum circuits, and show that they achieve nearly similar accuracies when compared to gradient-based methods. While sNES works pretty good for deep PQCs, xNES performs nearly well for shallow PQCs and outperforms sNES for problems where the parameters are correlated. We observe that the convergence curves for both NES and gradient-based methods are very similar, however, the number of circuit evaluations decrease significantly for NES and the initial rate of convergence seems to be higher too. 

We also extend the optimization problem to deep quantum circuits by using of batch optimization strategy and show that one can use NES with batch optimization for deep PQCs. Based on various numerical simulation we observe that we can increase the rate of convergence by choosing appropriate values of different hyper-parameters, learning rate, number of walkers, and batch size. We also show that random partition of parameters into batches works comparably well as structured partitioning. 

In the end we proposed a hybrid strategy, where we show that using NES we can amplify the gradients significantly and then use gradient-based methods for faster convergence. Our results indicate that the NES can be used as alternative to gradient-based methods in regions of vanishing gradients and can be a useful tool for exploring deep quantum circuits for different tasks. 

Our study only explores the basic use of NES for optimization of randomly initialized PQCs, we believe that they can be used with different gradient-based methods to improve the overall optimization of deep quantum circuits. Other possibilities like, using other distributions to sample the parameters from, or combination of strategies to partition parameters into batches and adjusting the hyper-parameters for the problem in hand, can be used to further improve the training. These different directions are something we leave for future studies, as they require additional research.

\section{Acknowledgement}
The authors thank Sukin Sim and Jakob Kottmann for valuable suggestions and comments on the manuscript.
A.A and A.A.G. acknowledge support by the  U.S. office of naval research. M.D. and A.A.G acknowledge support by the U.S. Department of Energy, Office of Science, Office of Advanced Scientific Computing Research, Quantum Algorithms Teams Program. A.A.G. acknowledges the generous support from Google, Inc. in the form of a Google Focused Award.  A.A.G. is grateful for the funding from the Vannevar Bush Faculty Fellowship Program, and the Canada Research Chairs Program. A.A.G thanks Anders G. Fr\o{}seth for his generous support. Simulations were performed on the Niagara supercomputer at the SciNet HPC Consortium~\cite{Loken_2010, ponce_2019}. SciNet is funded by: the Canada Foundation for Innovation; the Government of Ontario; Ontario Research Fund - Research Excellence; and the University of Toronto.

\section*{Data availability statement}
The data that support the findings of this study are available upon reasonable request from the authors.

\bibliography{QES.bib}

\end{document}